\documentclass[10pt,conference,letterpaper]{IEEEtran}
\usepackage{times,amsmath,epsfig}
\usepackage{graphicx}
\graphicspath{ {temp/} }
\usepackage{amsmath}
\usepackage{fdsymbol}
\usepackage{bbding}
\usepackage{fontawesome}
\usepackage{url}
\usepackage{subcaption}
\usepackage{listings}
\usepackage{booktabs} 
\usepackage[disable]{todonotes}

\usepackage{amsthm}

\theoremstyle{definition}
\newtheorem{definition}{Definition}

\newcommand{\para}[1]{\vspace{2mm}\noindent\textbf{#1}}

\title{Benchmarking Distributed Stream Data Processing Systems}
\author{%
{Jeyhun Karimov{\small $^{\#}$}, Tilmann Rabl{\small $^{*\#}$}, Asterios Katsifodimos{\small $^{\$}$}}
~\\
{Roman Samarev{\small $^{*}$}, Henri Heiskanen{\small $^{\ddag}$}, Volker Markl{\small $^{*\#}$} }
\vspace{1.6mm}\\
\fontsize{10}{10}\selectfont\itshape
$^{\#}$DFKI, Germany
$^{*}$TU Berlin, Germany
$^{\$}$Delft University of Technology, Netherlands
$^{\ddag}$Rovio Entertainment\\
}

 \everycr={\noalign{\global\advance\stepno by 1}}%
\usepackage{tikz}

\begin{document}
\linespread{0.955}
\maketitle
%
\begin{abstract}
	The need for scalable and efficient stream analysis has led to the development of many open-source streaming data processing systems (SDPSs) with highly diverging capabilities and performance characteristics. 
	While first initiatives try to compare the systems for simple workloads, there is a clear gap of detailed analyses of the systems' performance characteristics.
	In this paper, we propose a framework for benchmarking distributed stream processing engines. 
	We use our suite to evaluate the performance of three widely used SDPSs in detail, namely Apache Storm, Apache Spark, and Apache Flink. 
	Our evaluation focuses in particular on measuring the throughput and latency of windowed operations, which are the basic type of operations in stream analytics. 
	For this benchmark, we design workloads based on real-life, industrial use-cases inspired by the online gaming industry. 
	The contribution of our work is threefold.
	First, we give a definition of latency and throughput for stateful operators. 
	Second, we carefully separate the system under test and driver, in order to correctly represent the open world model of typical stream processing deployments and can, therefore, measure system performance under realistic conditions.
	Third, we build the first benchmarking framework to define and test the sustainable performance of  streaming systems.
	Our detailed evaluation highlights the individual characteristics and use-cases of each system.
\end{abstract}

%

\section{Introduction}
Processing large volumes of data in batch is often not sufficient in cases where new data has to be processed fast to quickly adapt and react to changes. 
For that reason, stream data processing  has gained significant attention. 
The most popular streaming engines, with large-scale adoption in industry and the research community, are Apache Storm \cite{toshniwal2014storm}, Apache Spark  \cite{zaharia2012discretized}, and Apache Flink \cite{carbone2015apache}. 
As a measure for popularity, we consider the engines' community size, pull requests, number of contributors, commit frequency at the source repositories, and the size of the industrial community adopting the respective systems in their production environment.

An  important application area of stream data processing is  online video games. 
These require the fast processing  of large scale online data feeds  from different sources. 
Windowed aggregations and windowed joins are two main operations that are used to monitor user feeds. 
A typical use-case is tracking the in-application-purchases per application, distribution channel, or product item (in-app products). 
Another typical use-case is the monitoring of advertising:  making sure that all campaigns and advertisement networks work flawlessly, and comparing different user feeds by  joining them. 
For example, monitoring the in-application-purchases of the same game downloaded from different distribution channels and comparing users'  actions  are essential in  online video game monitoring.

In this work, we propose a benchmarking framework to accurately measure the performance of SDPSs. 
For our experimental evaluation, we test three publicly available open source engines: Apache Storm, Apache Spark, and Apache Flink. 
We use latency and throughput as the two major performance indicators. 
Latency, in SDPS, is the time difference between the moment of data production at the source (e.g., the mobile device) and the  moment that the tuple has produced an output. 
Throughput, in this scenario, determines the number of ingested and processed records per time unit.

Even though there have been several comparisons of the performance of SDPS recently, they did not measure the latency and throughput that can be achieved in a production setting. 
One of the repeating issues in previous works is the missing definition and inaccurate measurement of latency in stateful operators (e.g., joins). 
Moreover, previous works do not clearly separate the system under test (SUT) and the benchmark driver. 
Frequently, the performance metrics are measured and calculated within the SUT resulting in incorrect measurements.

In this paper, we address the above mentioned challenges. 
Our proposed benchmarking framework is generic with a clear design and well defined metrics, which can be applied to a number of different stream execution engines. 
The main contributions of this paper are as follows:
\begin{itemize}
	\item We introduce a mechanism to accurately measure the latency of stateful operators in SDPSs. 
	We apply the proposed method to various use-cases. 
	\item We accomplish the complete separation of the test driver from the system under test. 
	\item We measure the maximum sustainable throughput of a SDPSs. 
	Our benchmarking framework handles system specific features like backpressure to measure the maximum sustainable throughput of a system. 
	\item We use the proposed benchmarking system for an extensive evaluation of Storm, Spark, and Flink with practical use-cases.
\end{itemize}

\section{Related work}
\label{rel}

Benchmarking parallel data processing systems has been an active area of research. Early benchmarking efforts have focused on batch processing and later on extended to stream processing.

\para{Batch Processing.}
HiBench~\cite{huang2011hibench} was the first benchmark suite to evaluate and characterize the performance of Hadoop and it was later extended with a streaming component~\cite{white2012hadoop}.
HiBench includes a wide range of experiments ranging from micro-benchmarks to machine learning algorithms. 
SparkBench, features machine learning, graph computation, SQL queries, and streaming applications on top of Apache Spark~\cite{li2015sparkbench}.
BigBench~\cite{ghazal2013bigbench} built an end-to-end benchmark with all major characteristics in the lifecycle of big data systems.
The BigDataBench~\cite{wang2014bigdatabench} suite contains  19 scenarios covering a broad range of applications and diverse data sets. Marcu et al.~\cite{marcu2016spark}
performed  an extensive analysis of the differences between Apache Spark and Apache Flink on iterative workloads. The above benchmarks either adopt batch processing systems and metrics used in batch processing systems or apply the batch-based metrics on SDPSs. We, on the other hand, analyze streaming systems with a new definition of metrics and show that adopting batch   processing metrics for SDPSs leads to biased benchmark results.

\para{Stream Processing.} 
Recently, a team from Yahoo! conducted an informal series of experiments on three Apache projects, namely Storm, Flink, and Spark and measured their latency and throughput \cite{chintapalli2016benchmarking}. 
They used Apache Kafka \cite{kreps2011kafka} and Redis \cite{carlson2013redis} for data retrieval and storage respectively. 
Perera et al. used the Yahoo Streaming Benchmark and Karamel \cite{karamel} to provide reproducible batch and streaming benchmarks of Apache Spark and Apache Flink in a cloud environment \cite{perera2016reproducible}. 
Later on, it was shown \cite{dataartisans} that Kafka and Redis were the bottleneck in the experiments of the Yahoo! Streaming Benchmark \cite{chintapalli2016benchmarking} and transitively to \cite{perera2016reproducible}. 
In this paper, we overcome those  bottlenecks by $i)$~generating the data on the fly with a scalable data generator (Section \ref{des}) instead of ingesting data from Kafka and $ii)$~not storing data in a key-value store.

Lopez et al.~\cite{lopez2016performance} propose a benchmarking framework to assess the throughput performance of Apache Storm, Spark, and Flink under node failures. 
The key finding of their work is that Spark is more robust to node failures but it performs up to an order of magnitude worse than Storm and Flink. 
Compared to this work, we observed a large difference with respect to the throughput achieved by the same systems. 
The paper in question allows the systems to ingest data at maximum rate. 
Instead, we introduce the concept of sustainable throughput: in our experiments we control the data ingestion rate (throughput) of a system, in order to avoid large latency fluctuations. 
We argue that  sustainable throughput is a more representative metric which takes into account the latency of a system.

Shukla et al.~\cite{shukla2016benchmarking} perform common IoT tasks with different SDPSs, evaluating their performance. 
The authors define latency as the interval between the source operator's \emph{ingestion} time and the sink operator's result \emph{emission} time. 
As we discuss in Section~\ref{sec_metrics}, this approach leads to inaccurate measurements as the effects of backpressure and other effects are not visible in the results. The same issue is also present in the LinearRoad benchmark~\cite{arasu2004linear}. To alleviate this problem, we perform experiments measuring \emph{event-time latency}. 
Additionally, Shukla et al. define throughput  as the  rate of \emph{output} messages emitted from the output operators in a unit time. 
However, since the number of result-tuples can differ from the input-tuples (e.g., in an aggregation query) we measure the throughput of \emph{data ingestion} and introduce the concept of sustainable throughput.

StreamBench~\cite{lu2014stream}, proposes a method to measure the throughput and latency of a SDPS with the use of a mediator system between the data source and the SUT. 
In this work, we explain that such a mediator system is  a bottleneck and/or affect the measurements' accuracy. 
Finally, several stream processing systems implement their own benchmarks to measure the system performance without comparing them with any other system~\cite{neumeyer2010s4,qian2013timestream,zaharia2012discretized}. 

In summary, our benchmark framework is the first to $i)$~separate the SUT and driver, $ii)$~use a scalable data generator and to $iii)$~define metrics for system-, and event-time, as well as to $iv)$~introduce and use the concept of sustainable throughput throughout experiments.

\section{Benchmark Design Decisions}
\label{des}
In this section, we  discuss the main design decisions of our benchmarking framework. We choose to generate data on-the-fly, rather than reading the data from a message broker or the filesystem and we use queues between the data generators and the streaming systems.


\subsection{On-the-fly Data Generation vs. Message Brokers}
Streaming systems nowadays typically pull the data from message brokers, such as Apache Kafka~\cite{kreps2011kafka}, instead of directly connecting to push-based data sources. 
The message broker persists data coming from various sources~\cite{ranjan2014streaming}, allowing for data replication and making it available for other systems to use. 
The data exchange between the message broker and the streaming system may easily become the bottleneck of a benchmark deployment for a number of reasons. 
First, if the message broker's data partitioning is not chosen wisely, data re-partitioning may occur before the data reaches the sources of the streaming system.
This can happen when data resides in a different machine in the cluster or the data is partitioned in a different way than the streaming system requires it. 
Finally, the data needs to persist on disk before going through a de-/serialization layer between the streaming system and the message broker.
In our benchmark design, we choose to not use  a message broker, but rather, use a distributed in-memory data generator with configurable data generation rate. Before each experiment we benchmarked and distributed our data generator such that the data generation rate is faster than the data ingestion rate of the fastest system. 
This way, the communication between the data generator and the SUT is bounded only by the network bandwidth and the speed of the data ingestion by the SUT.

\subsection{Queues Between Data Generators and SUT Sources}
It is quite common that the data ingestion rate or throughput of a streaming system is not constant throughout the duration of an experiment. The fluctuations in the ingestion rate can be due to transient network issues, garbage collection in JVM-based engines, etc.
To alleviate this problem, we add a queue between each data generator and the SUT's source operators in order to even out the difference in the rates of data generation and data ingestion.

\subsection{Separation of Driver and the SUT}
We choose to isolate the benchmark driver, i.e., the data generator, queues, and measurements from the SUT. 
In previous works, the throughput was either measured inside the SUT or the benchmark used internal statistics of the SUT.
However, different systems can have very diverse definitions  of latency and throughput leading.
In our benchmarking framework, we choose to separate the driver and the SUT, to perform measurements out of the SUT. 
More specifically, we measure throughput at the queues between the data generator and the SUT and measure latency at the sink operator of the SUT. 
Each pair of data generator and queue resides on the same machine to avoid any network overhead and to ensure data locality, while the queue data is always kept in memory to avoid disk write/read overhead.

The data generator timestamps each event at generation time. It performs so, with constant speed throughout the experiment. 
The event's latency is calculated from the time instance that it is generated, i.e., the longer an event stays in a queue, the higher its  latency. 
We make sure that no driver instance runs on the same machines as the SUT  to affect its performance.

\begin{figure*}[!htb]
	\centering
	\includegraphics[width=0.9\textwidth]{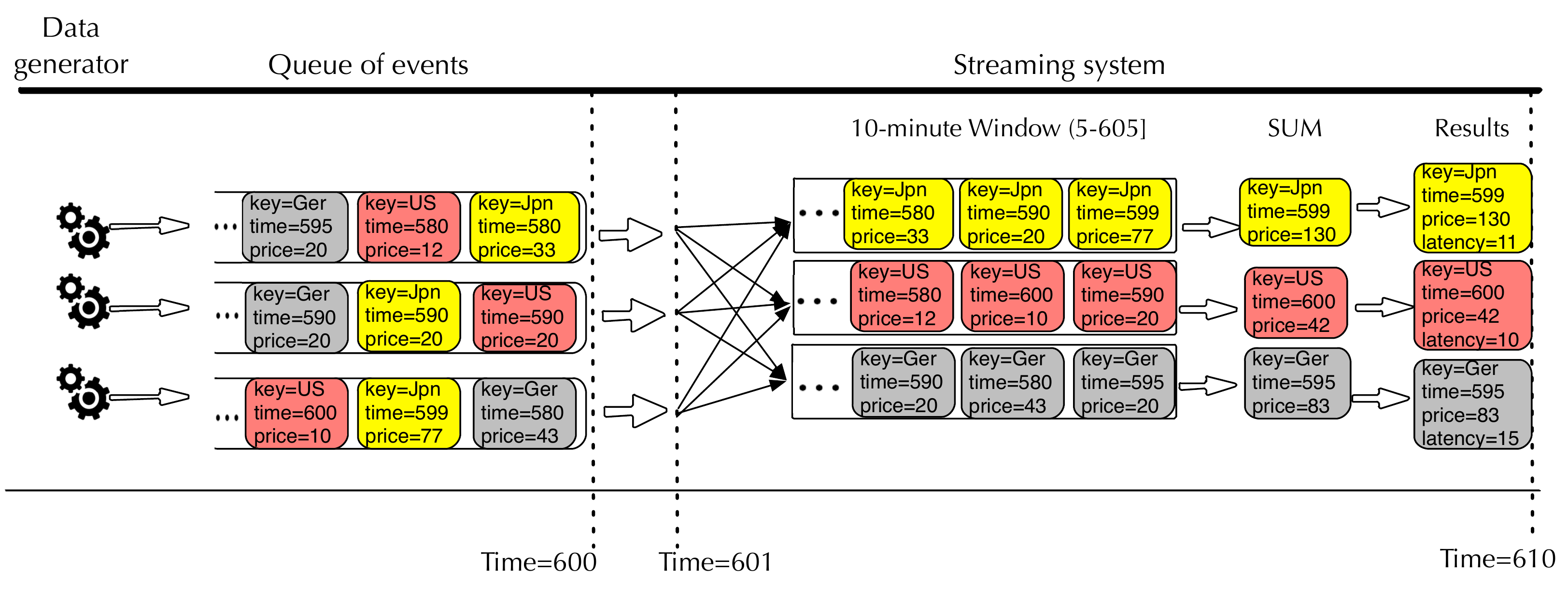}
		
	\caption{End-to-end example of an aggregation query. The data generator produces tuples on a given event-time (before \texttt{time=600}), a second later the events enter the streaming system where they are grouped by their key, and finally aggregated (SUM). The event-time latency of the output tuple equals to the maximum event-time latency of elements in each window.}
	\label{agg_usecase}
\end{figure*}

\begin{figure*}[!htb]
	\centering
	\includegraphics[width=0.9\textwidth]{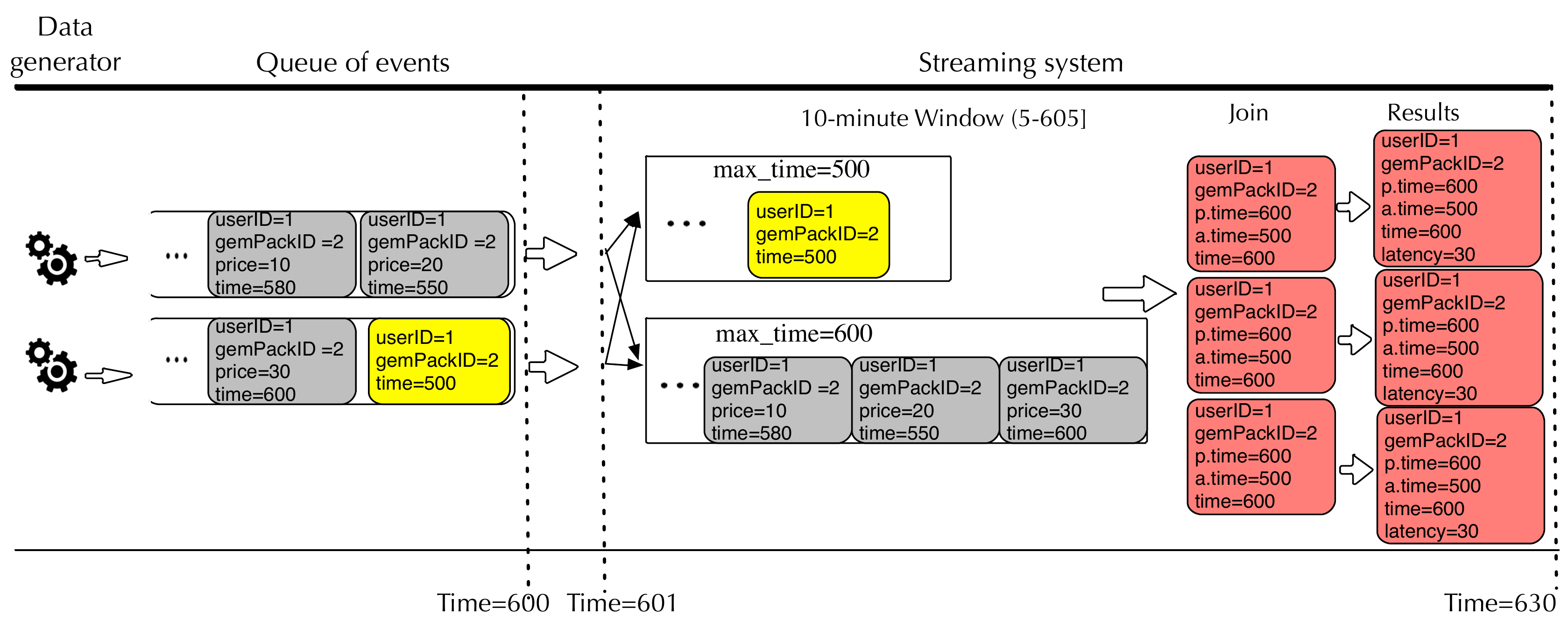}
		
	\caption{End-to-end join of two streams. The generator creates events, which are then read by the streaming system and form a 10-minute window. The events are joined and the event-time of the result-tuples equals to the maximum event-time  of tuples in their corresponding windows.}
	\label{join_usecase}
\end{figure*}

\section{Metrics}
\label{sec_metrics}

Streaming systems are typically evaluated using two main metrics: throughput and latency. In this section, we make a distinction between two types of latency, namely event-time latency and processing-time latency. We then describe two types of throughput, namely maximum throughput and sustainable throughput. 

\subsection{Latency}

Modern streaming systems~\cite{akidau2015dataflow,carbone2015apache,akidau2013millwheel}  distinguish two notions of time: \textit{event-time} and \textit{processing-time}. 
The \textit{event-time} is the time when an event is captured while \textit{processing-time} is the time when an operator processes a tuple. Similar to the nomenclature of these two notions of time, we distinguish between event- and processing-time latency.

\begin{definition}[Event-time Latency]
We define \emph{event-time latency} to be the interval between a tuple's event-time and its emission time from the SUT output operator. 
\end{definition}

For instance in an ATM transaction, the event-time is the moment of a user's action at the terminal, and the event-time latency is the time interval between the moment that the user's action took place and the moment that the event has been fully processed by the streaming system. 

\begin{definition}[Processing-time Latency]
We define \emph{processing-time latency} to be the interval between a tuple's \emph{ingestion} time (i.e., the time that the event has reached the input operator of the streaming system) and its emission time from the SUT output operator.  
\end{definition}

\para{Event- vs. processing-time latency.} Event- and processing-time latencies are equally important metrics. The event-time latency includes the time that a given event has spent in a queue, waiting to be processed, while processing-time latency is used to measure the time it took for the event to be processed by the streaming system. In practical scenarios, event-time latency is very important as it defines the time in which the user interacts with a given system and should be minimized. Clearly, processing-time latency makes part of the event-time latency. We use both metrics to characterize a system's performance.
Not clearly differentiating the two metrics leads to the coordinated omission problem. In coordinated omission service time (i.e., processing time) is measured at the SUT and any increasing queuing time, which is part of the \textit{response time} (i.e., event time), is ignored \cite{omission}. 
Friedrich et al. show that coordinated omission leads to significant underestimation of latencies \cite{friedrich2017coordinated}.

\para{Event-time Latency in Windowed Operators.} 
Stateful operators such as window aggregates (a sum aggregate over an hour's worth of data), retain state and return results after having seen a number of tuples over some time.
Measuring latency in such cases is non-trivial. 
The reason is that the latency of a given windowed operator is affected by the tuples' waiting time until the window is formed completely.

Figure \ref{agg_usecase} depicts the data generator, and a set of events in three queues. The events are timestamped with their event-time when they are generated. The events are then grouped by their key and are put in a 10-minute window.

Take, for example, the window containing the red events with \texttt{key=US}. The timestamps of these three events are of 580, 590, and 600. When these events are aggregated into a new event (the sum of their values with a total of \texttt{value=42}) we need to assign an event-time to that output. That event-time is then used to calculate the event-time latency on the operator output (in this case, \texttt{latency=10}). The main intuition is that in this way, we exclude the tuples' waiting time while the window is still buffering data. The event-time is defined more formally below.

\begin{definition}[Event-time of Windowed Events]
The event-time of a windowed operator's output event,  is the \emph{maximum event-time of all events that contributed to that output}.  
\end{definition}

In a windowed join operation, the containing tuples' event-time is set of be the maximum event-time of their window. Afterwards, each join output is assigned the maximum event-time of its matching tuples. As described in our example, in order to calculate the event-time latency of an output tuple, all we have to do is subtract the event-time of that tuple from the current system time. Figure \ref{join_usecase} shows the main intuition behind this idea. We join ads (yellow) and purchases (green) streams in a 10-minute window.

\para{Processing-time Latency in Windowed Operators.} 
Apart from event-time latency, we need to calculate the processing-time latency of tuples as well. We define the processing-time of a windowed event similarly to the event-time.

\begin{definition}[Processing-time of Windowed Events]
The processing-time of a windowed operator's output event, is the \emph{maximum processing-time of all events that contributed to that output}.  
\end{definition}

The processing-time latency is calculated in exactly the same way as for event-time, with a small difference. Every tuple is enriched with an extra, processing-time field at its ingestion time (when it reached the first operator of the streaming system). In our example in Figure \ref{agg_usecase}, this happens right after \texttt{time=601}. To calculate the processing-time latency, we simply subtract the processing-time of that tuple from the current system time.


\subsection{Throughput}
The throughput of a data processing system is defined as the number of events that the system can process in a given amount of time. Throughput and event-time latency often do not correlate. For instance, a streaming system that batches tuples together before processing them, can generally achieve higher throughput. However, the time spent batching events affects the events' event-time latency. 

In practice, the deployment of a streaming system has to take into account the arrival rate of data. When the data arrival rate increases, the system has to adapt (e.g., by scaling out) in order to  handle the increased arrival rate and process tuples without exhibiting backpressure. To reflect this, we define the concept of \emph{sustainable throughput} and discuss how we attain it in our experiments.

\begin{figure}[!htb]
	\centering
	\includegraphics[width=0.5\textwidth]{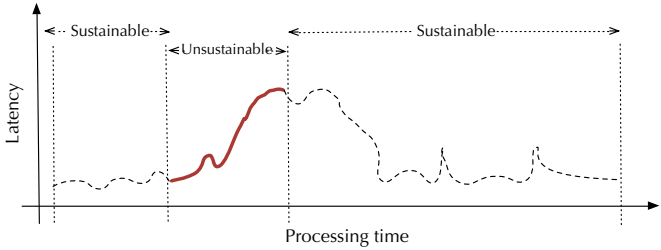}
		
	\caption{Sustainable throughput and unsustainable throughput}
	\label{fig:sustainable-throughput}
\end{figure}

\para{Sustainable Throughput.} When given more data than it can handle, a system starts to build up backpressure, i.e., the system queues up new events in order to process the events that have already been ingested. As a result, from the moment that backpressure mechanism is started, the event-time latency of all queued events increases. As we can see from Figure \ref{fig:sustainable-throughput},  backpressure can be transient: as soon as the system catches up again with the events' arrival rate, the event-time latency will stabilize. When the system's throughput is larger than the events' arrival rate, the event-time latency will decrease to the minimum (i.e., the processing-time).

\begin{definition}[Sustainable Throughput]
Sustainable throughput is the highest load of event traffic that a system can handle without exhibiting prolonged backpressure, i.e., without a continuously increasing event-time latency.
\end{definition}

In our experiments, we make sure that the data generation rate matches the sustainable throughput of a given deployment. To find the sustainable throughput of a given deployment we run each of the systems with a very high generation rate and we decrease it until the system can sustain that data generation rate. We allow for some fluctuation, i.e., we allow a maximum number of events to be queued, as soon as the queue does not continuously increase. 

\section{Workload Design}
The workload for our benchmark is derived from an online video game application at Rovio\footnote{Creator of the Angry Birds game: \url{http://www.rovio.com/}.}. Rovio continuously monitors the user actions of a given a game to ensure that their services work as expected. 
For instance, they continuously monitor the number of active users and generate alerts when this number has large drops.
Moreover, once a game has an update or receives a new feature, Rovio monitors incoming events to check whether the newly added feature is working without issues.
Rovio also tracks the in-app purchases per game, and distribution channel (e.g., Apple's AppStore, Google Play), and per in-app purchased item (e.g., a gem pack) and proposes gem packs to users as their game progresses.

\para{Dataset.}
We focus on the advertisements of gem packs. As seen in Listing 1, we have two data streams $i)$~the \texttt{purchases} stream, which contains tuples of purchased gem packs along with the purchase time, the user that made the purchase and the gem pack that was bought and $ii)$~the \texttt{ads} stream which contains a stream of proposals of gem packs to users, at given time instants. We used synthetic data, which we generate with our data generator.

\para{Queries.}
The first query that we use is an aggregation query. More specifically, we need to have a sliding window of revenue made from each gem pack. The template we use to generate such queries can be found in Listing 1.

The second query that we use is a typical use-case of correlating advertisements with their revenue. As we can see in the bottom of Listing 1, each user is presented with a specified proposal to buy a gem pack at a given time instant. We join this stream with the stream of purchases in order to find which of the proposed gems has been bought, as a result of proposing the gem to users.


\begin{figure}
\begin{lstlisting}[
language=SQL,
showspaces=false,
basicstyle=\small,
commentstyle=\color{gray}
label=list2,
caption={Query templates used by our workloads.},
captionpos=b]

# Streams
PURCHASES(userID, gemPackID, price, time)
ADS(userID, gemPackID, time)

# Windowed Aggregation Query
SELECT SUM(price)
FROM   PURCHASES [Range r, Slide s] 
GROUP  BY gemPackID

# Windowed Join Query
SELECT p.userID, p.gemPackID, p.price
FROM   PURCHASES [Range r, Slide s] as p, 
       ADS       [Range r, Slide s] as a, 
WHERE  p.userID = a.userID AND
       p.gemPackID = a.gemPackID
\end{lstlisting}
\end{figure}
\section{Evaluation}
\label{eval}
In this section, we evaluate the performance of three streaming systems, namely Storm 1.0.2, Spark 2.0.1, and Flink 1.1.3.
Due to the large number of parameters and variables, it is not possible not include all the experimental results in this paper. 
Instead, we present the most interesting results.

\subsection{System setup}
Our cluster consists of 20 nodes, each equipped with 2.40GHz Intel(R) Xeon(R) CPU  E5620  with 16 cores and 16GB RAM. The network bandwidth is 1Gb/s. We use a dedicated master for the streaming systems and an equal number of workers and driver nodes (2, 4, and 8). 
All nodes' system clocks in the cluster are synchronized via a local NTP server. Each data generator generates 100M events with constant speed using 16 parallel instances.
We generate events with normal distribution on key field.


We use 25\% of the input data as a warmup. 
We enable  backpressure  in all systems, that is, we do not allow the systems to ingest more input than they can process and crash during the experiment. 
If the SUT drops one or more connections to the data generator queue, then the driver halts the experiment with the conclusion that the SUT cannot sustain the given throughput. 
Similarly, in real-life if the system cannot sustain the user feed and drops connection,  this is considered as a failure.

Throughout the experiments, if the behavior of the system is similar with different parameters, we select the most interesting figures or features due to the limited space.

\para{Tuning the systems.} \label{sec:tuning}
Tuning the engines' configuration parameters is important to get a good performance for every system.
We adjust the buffer size  to ensure a good balance between throughput and latency.
Although selecting low buffer size can result in a low processing-time latency, the event-time latency of tuples may increase as they will be queued in the driver queues instead of the buffers inside the streaming system. 
Moreover, we adjust the block interval for partitioning the RDDs in Spark.  
The number of RDD partitions a single mini-batch  is bounded by $\frac{batch \ Interval}{block \ Interval}$. 
As the cluster size increases, decreasing the block interval can increase the parallelism. 
One of the main reasons that Spark scales up very well is the partitioning of RDDs. 
However, depending on the use-case, the optimal number of RDD partitions can change. 
In Storm the number of workers, executors, and buffer size are the  configurations (among many other) that need to be tuned to get the best performance. 
For all systems,  choosing the right level or parallelism is essential to balance between good resource utilization and network or resource exhaustion.

Storm introduced the  backpressure feature in recent releases; however, it is not mature yet. 
With high workloads, it is possible that the backpressure stalls the topology, causing spouts to stop emitting tuples. 
Moreover, we notice that Storm  drops some connections to the data queue when tested with high workloads with backpressure disabled, which is not acceptable according to the real world use-cases. 
Dropping connections due to high throughput is considered  a system failure.

\subsection{Performance Evaluation}

    \begin{table}
    \centering
        \begin{tabular}{llllll}\toprule
            &\textbf{2-node}  & \textbf{4-node} & \textbf{8-node}\\\midrule
            Storm & 0.4 M/s & 0.69 M/s & 0.99 M/s \\
            Spark & 0.38 M/s & 0.64 M/s & 0.91 M/s \\
            Flink & 1.2 M/s & 1.2 M/s & 1.2 M/s\\
        \end{tabular}
        \caption{Sustainable throughput for windowed aggregations}
        \vspace{-2em}
        \label{tab_th_agg}
    \end{table}

    \begin{table*}
	\resizebox{\textwidth}{!} {\begin{tabular}{ c |c c c c | c c c c | c c c c}\toprule
			& & & \textbf{2-node}  & & & & \textbf{4-node} & & & & \textbf{8-node}\\
			& \textit{avg} & \textit{min} & \textit{max} & \textit{quantiles (90,95,99)} & \textit{avg} & \textit{min} & \textit{max} & \textit{quantiles (90,95,99)} & \textit{avg} & \textit{min} & \textit{max} & \textit{quantiles (90,95,99)} \\\midrule
			Storm & 1.4 &0.07 &  5.7 & (2.3, 2.7, 3.4) & 2.1 & 0.1 & 12.2 & (3.7, 5.8, 7.7) & 2.2 & 0.2 & 17.7 & (3.8, 6.4, 9.2) \\
			Storm(90\%) & 1.1 & 0.08 & 5.7 & (1.8, 2.1, 2.8) & 1.6 & 0.04 & 9.2 & (2.9, 4.1, 6.3) & 1.9 & 0.2 & 11 & (3.3, 5, 7.6)\\
			Spark & 3.6 & 2.5 & 8.5 & (4.6, 4.9, 5.9) & 3.3 & 1.9 & 6.9 & (4.1, 4.3, 4.9) & 3.1 & 1.2 & 6.9 & (3.8, 4.1, 4.7)\\
			Spark(90\%) & 3.4 & 2.3 & 8 & (3.9, 4.5, 5.4) & 2.8 & 1.6 & 6.9 & (3.4, 3.7, 4.8) & 2.7 & 1.7 & 5.9 & (3.6, 3.9, 4.8)\\
			Flink & 0.5 & 0.004 & 12.3 & (1.4, 2.2, 5.2) & 0.2 & 0.004 & 5.1 & (0.6, 1.2, 2.4) & 0.2 & 0.004 & 5.4 & (0.6, 1.2, 3.9)\\
			Flink(90\%) & 0.3 & 0.003 & 5.8 & (0.7, 1.1, 2) & 0.2 & 0.004 & 5.1 & (0.6, 1.3, 2.4)  & 0.2 & 0.002 & 5.4 & (0.5, 0.8, 3.4)\\
	\end{tabular}}
	\caption{Latency statistics, avg, min, max, and quantiles (90, 95, 99) in seconds for windowed aggregations}
	\vspace{-2em}
	\label{tab_lat_agg}
\end{table*}

\begin{figure*}
    \centering
    \begin{subfigure}[b]{0.28\textwidth}
        \includegraphics[width=\textwidth]{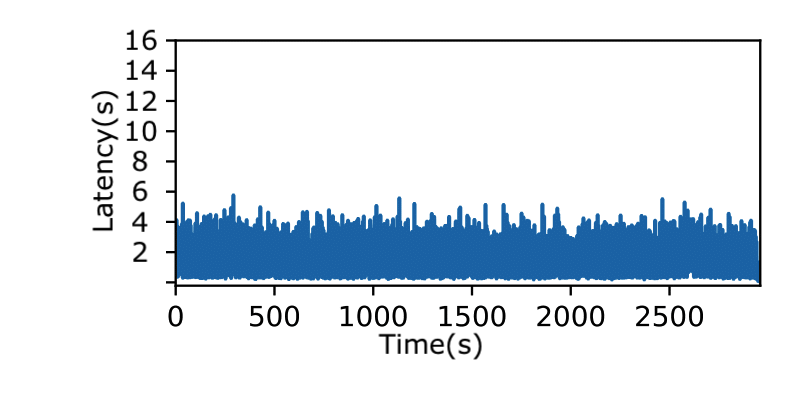}

        \caption{Storm, 2-node,  max   throughput }
    \end{subfigure}\hspace{-2mm}%
    \begin{subfigure}[b]{0.28\textwidth}
        \includegraphics[width=\textwidth]{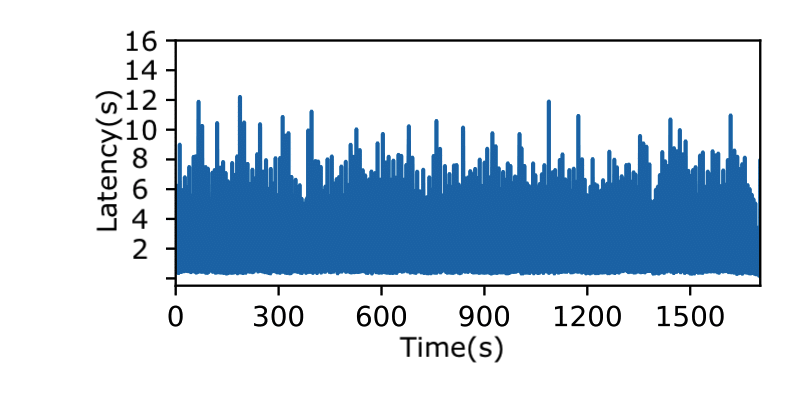}

        \caption{Storm, 4-node,  max   throughput }
    \end{subfigure}\hspace{-2mm}%
    \begin{subfigure}[b]{0.28\textwidth}
        \includegraphics[width=\textwidth]{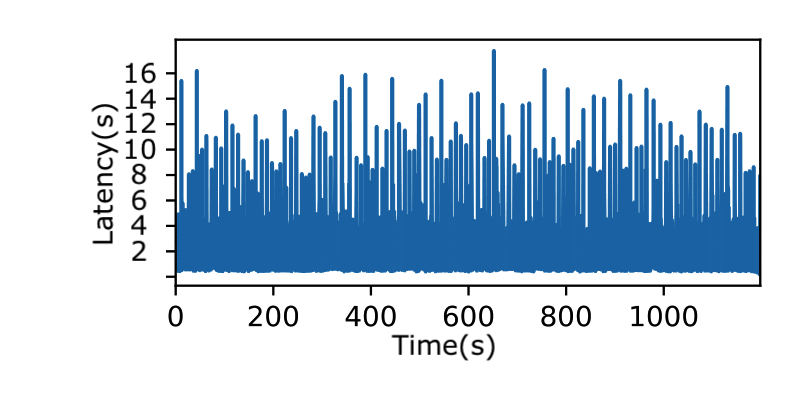}

        \caption{Storm, 8-node,  max   throughput }
                 \label{fig_storm_agg_8node_th_max_ts}
    \end{subfigure}

    \begin{subfigure}[b]{0.28\textwidth}
        \includegraphics[width=\textwidth]{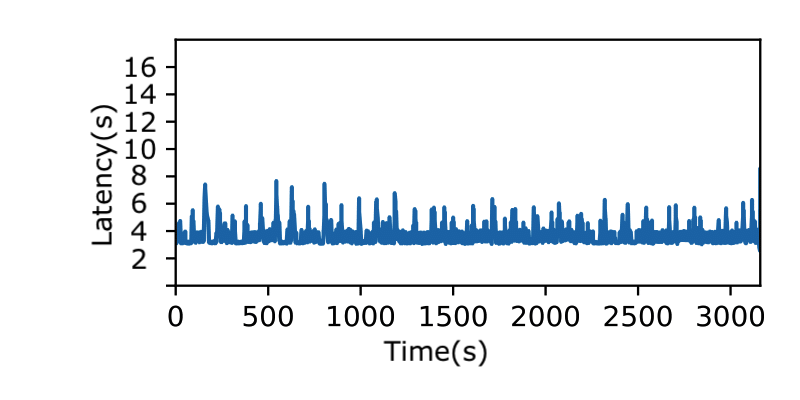}

        \caption{Spark, 2-node,  max   throughput }
    \end{subfigure}\hspace{-2mm}%
    \begin{subfigure}[b]{0.28\textwidth}
        \includegraphics[width=\textwidth]{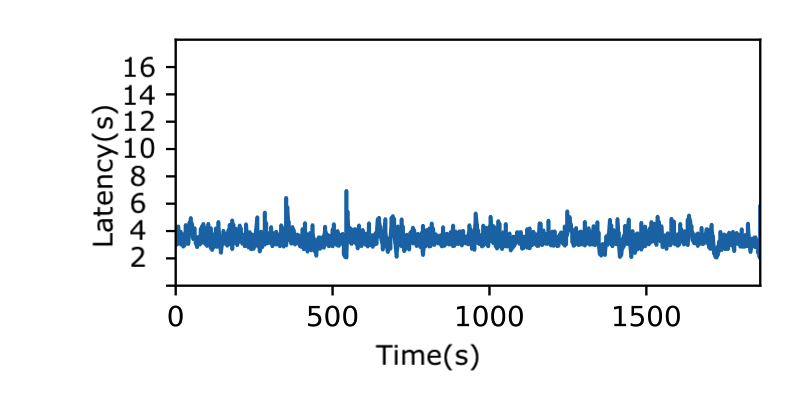}

        \caption{Spark, 4-node,  max   throughput }
    \end{subfigure}\hspace{-2mm}%
    \begin{subfigure}[b]{0.28\textwidth}
        \includegraphics[width=\textwidth]{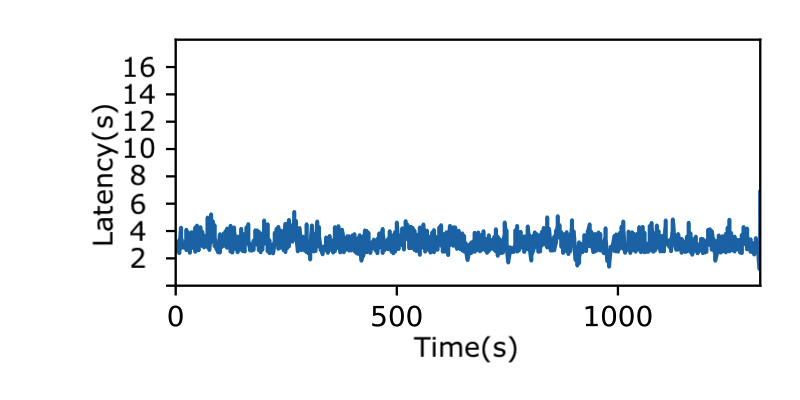}

        \caption{Spark, 8-node,  max   throughput }
         \label{fig_spark_agg_8node_th_max_ts}
    \end{subfigure}

    \begin{subfigure}[b]{0.28\textwidth}
        \includegraphics[width=\textwidth]{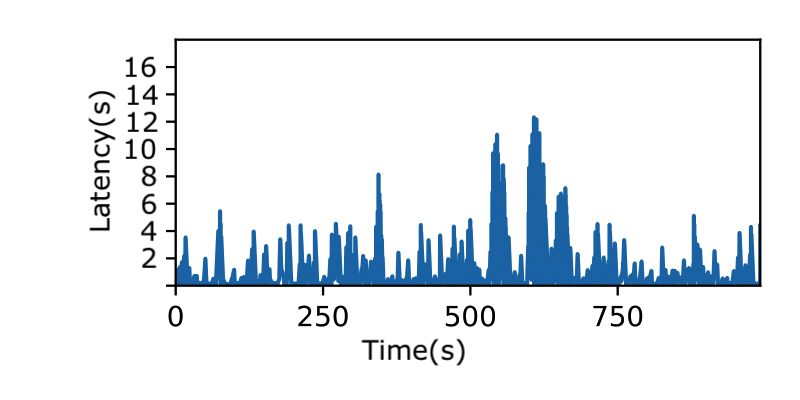}

        \caption{Flink, 2-node,  max   throughput }
                \label{flink_agg_2node_th_max_ts}

    \end{subfigure}\hspace{-2mm}%
    \begin{subfigure}[b]{0.28\textwidth}
        \includegraphics[width=\textwidth]{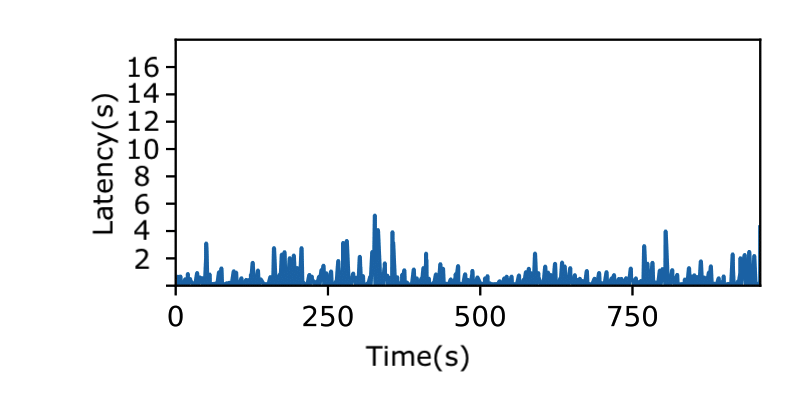}

        \caption{Flink, 4-node,  max   throughput }
    \end{subfigure}\hspace{-2mm}%
    \begin{subfigure}[b]{0.28\textwidth}
        \includegraphics[width=\textwidth]{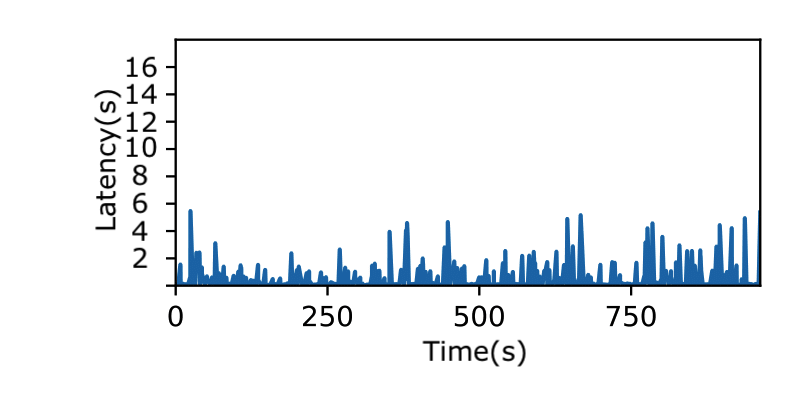}

        \caption{Flink, 8-node,  max   throughput }
        
    \end{subfigure}

    \begin{subfigure}[b]{0.28\textwidth}
        \includegraphics[width=\textwidth]{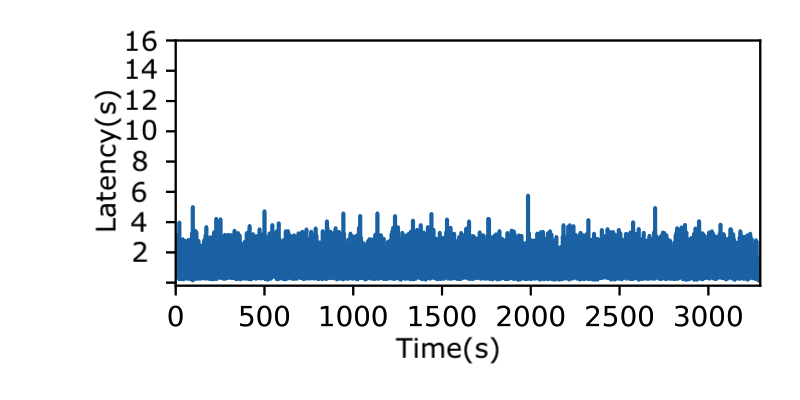}

        \caption{Storm, 2-node,   90\%- throughput }
    \end{subfigure}\hspace{-2mm}%
    \begin{subfigure}[b]{0.28\textwidth}
        \includegraphics[width=\textwidth]{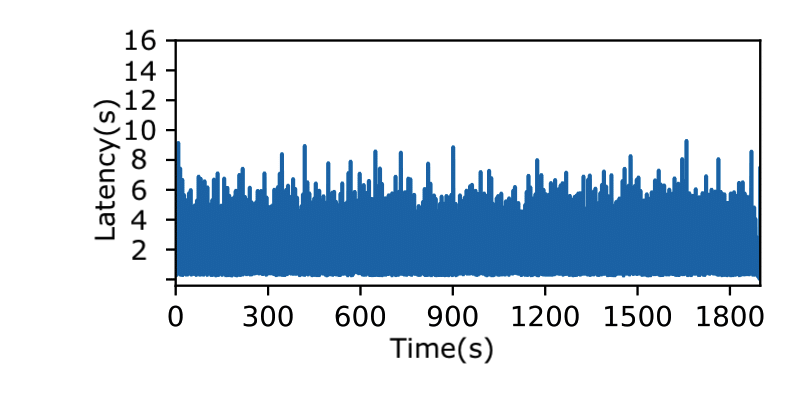}

        \caption{Storm, 4-node,  90\%- throughput }
                        \label{fig_storm_agg_8node_th_max_ts}
    \end{subfigure}\hspace{-2mm}%
    \begin{subfigure}[b]{0.28\textwidth}
        \includegraphics[width=\textwidth]{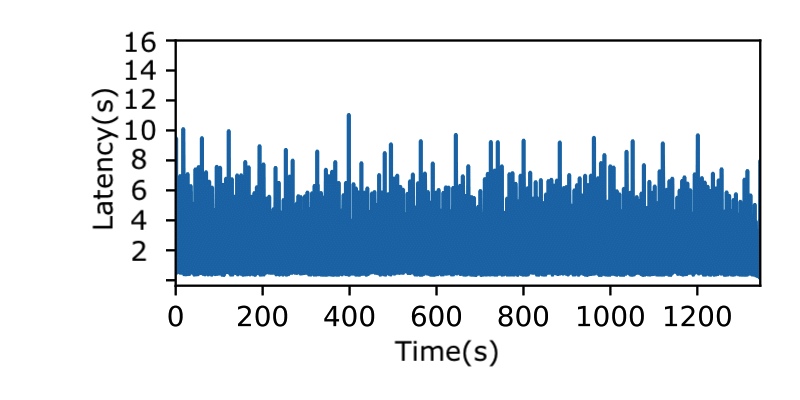}

        \caption{Storm, 8-node,  90\%- throughput }
                \label{fig_storm_agg_8node_th_90_ts}
    \end{subfigure}

    \begin{subfigure}[b]{0.28\textwidth}
        \includegraphics[width=\textwidth]{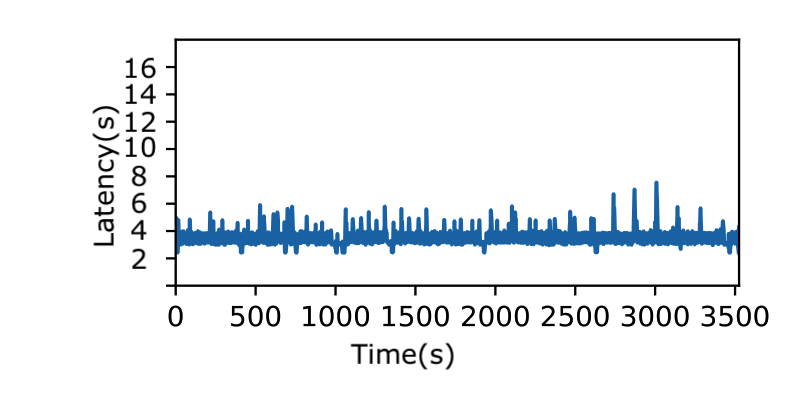}

        \caption{Spark, 2-node,   90\%- throughput }
    \end{subfigure}\hspace{-2mm}%
    \begin{subfigure}[b]{0.28\textwidth}
        \includegraphics[width=\textwidth]{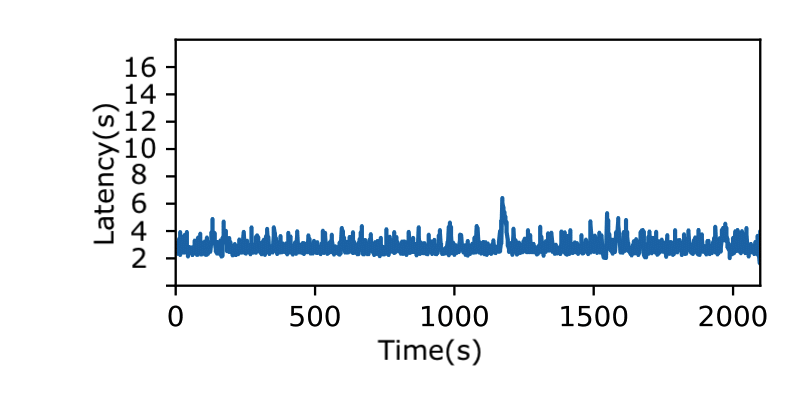}

        \caption{Spark, 4-node,   90\%- throughput }
    \end{subfigure}\hspace{-2mm}%
    \begin{subfigure}[b]{0.28\textwidth}
        \includegraphics[width=\textwidth]{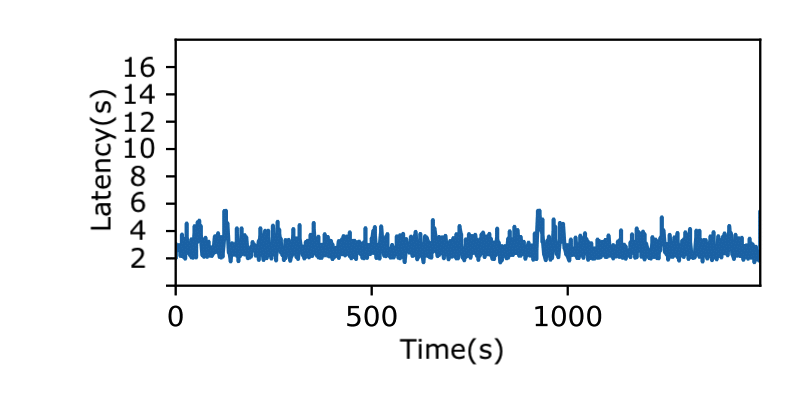}

        \caption{Spark, 8-node,   90\%- throughput }
        
    \end{subfigure}

    \begin{subfigure}[b]{0.28\textwidth}
        \includegraphics[width=\textwidth]{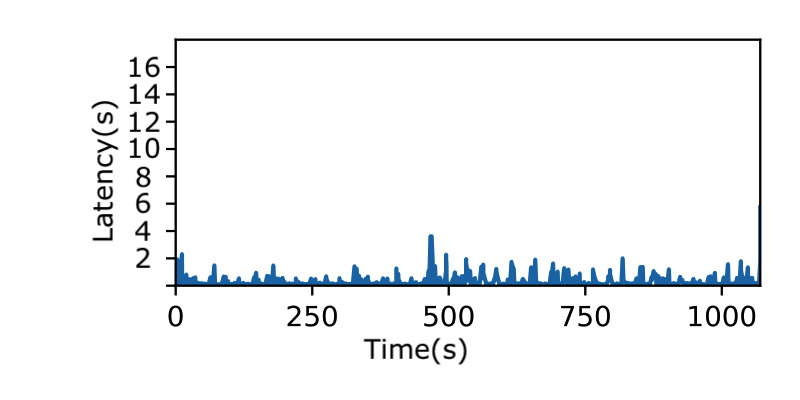}

        \caption{Flink, 2-node,   90\%- throughput }
        \label{fig_flink_agg_2node_th_90_ts}
    \end{subfigure}\hspace{-2mm}%
    \begin{subfigure}[b]{0.28\textwidth}
        \includegraphics[width=\textwidth]{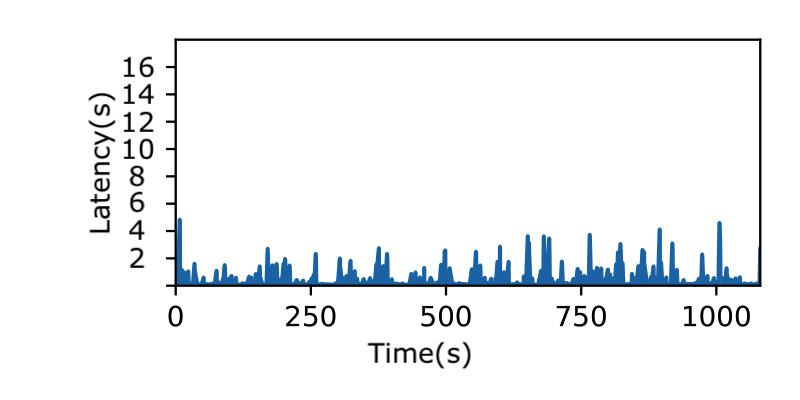}

        \caption{Flink, 4-node,   90\%- throughput }
    \end{subfigure}\hspace{-2mm}%
    \begin{subfigure}[b]{0.28\textwidth}
        \includegraphics[width=\textwidth]{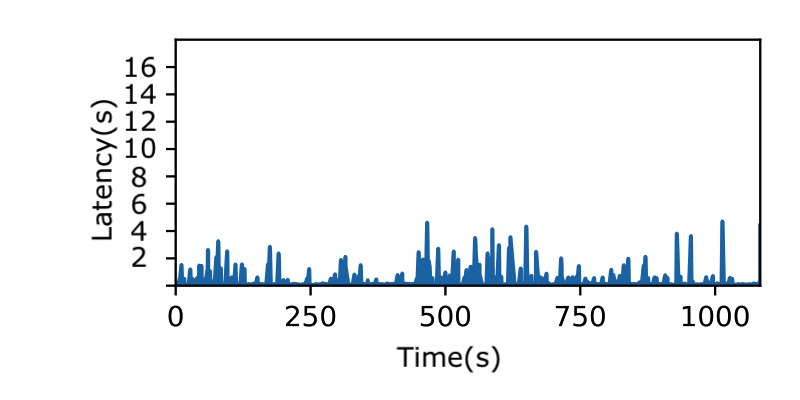}

        \caption{Flink, 8-node,  90\% throughput }
        
    \end{subfigure}

        \caption{Windowed aggregation latency distributions in time series}
        \vspace{-1em}
                \label{fig_ts_agg}
\end{figure*}

\para{Experiment 1: Windowed Aggregations.}
We use an  aggregation query (8s, 4s), 8 seconds window length and 4 seconds window slide, for our first evaluations.
 Table \ref{tab_th_agg} shows the sustainable throughput of the SDPSs. 
We use a four second batch-size for  Spark, as it can sustain the  maximum throughput with this configuration. 
We identify that Flink's performance is bounded by network bandwidth with    4- or more node cluster configuration. 
Storm's and Spark's performance in terms of throughput are comparable, with Storm outperforming Spark by approximately 8\% in all configurations.

Table \ref{tab_lat_agg} shows the latency measurements of windowed aggregations. 
We conduct experiments with maximum and 90\%-workloads. 
The latencies shown in this table correspond to    the workloads given in Table \ref{tab_th_agg}. 
In most cases, where the network bandwidth is not a bottleneck, we can see a significant decrease in latency when lowering the throughput by 10\%. 
This shows that the sustainable throughput saturates the system.


As we see from Table \ref{tab_lat_agg},  Spark has a higher latency than Storm and Flink but it exhibits less variation in $avg$, $min$, and $max$ latency measurements. 
Because Spark processes tuples in mini-batches, the tuples within the same batch have similar latencies and, therefore, there is little difference among the measurements. 
Moreover, transferring data from Spark's block manager to DStream by creating RDDs adds additional overhead that results in higher $avg$ latencies for Spark compared to Flink and Storm. 

The $avg$ and $max$ latencies increase in Storm with increasing workload and cluster size, while in Spark we see the opposite behavior, which means Spark can partition the data (RDDs) better in a bigger distributed environments. 
However, from the quantile values we can conclude that the $max$ latencies of Storm  can be considered as outliers.

Figure \ref{fig_ts_agg} shows the windowed aggregation latency distributions over time.
In all cases, we can see that the fluctuations are lowered when decreasing the throughput by 10\%. 
While in Storm and in Flink it is hard to detect the lower bounds of latency as they are close to zero, in Spark the upper and lower boundaries are more stable and clearly noticeable. 
The reason is that a Spark job's characteristics are  highly dependent on the batch size and this determines the clear upper and lower boundaries for the latency. 
The smaller the batch size, the lower the latency and throughput. 
To have a stable and efficient configuration in Spark, the mini-batch processing time should be less than the batch interval. 
We determine the most fluctuating system to be Flink in  2-node setup and Storm in 8-node setup  as shown in Figures \ref{flink_agg_2node_th_max_ts} and \ref{fig_storm_agg_8node_th_max_ts}. 
Those fluctuations show the behaviour of backpressure. 

Spark has a mini-batch based architecture and splits the input query into multiple sub-queries, which get executed in separate jobs. 
To analyze this from the logical query plan perspective,  each RDD has a $transformation()$ method. 
Each $transformation()$ method produces one or more RDDs of any type that can be expressed in Scala. 
Moreover, each $transformation()$ may contain several $sub$-$transformations$. 
As a result, coordination and pipelining mini-batch jobs and their stages creates extra overhead for Spark as the relationship between consequent RDDs in a job pipeline is not necessarily one-to-one. 
For example,  we use $reduceByKey()$ to parallelize the stages of the mini-batch within a  window. 
It is transformed into two subsequent RDDs: first a ShuffledRDD and then a MapPartitionsRDD. 
In Flink and Storm, on the other hand, this is just single step both in  logical and  physical query plan.

    \begin{table}
    \centering
        \begin{tabular}{lllll}\toprule
            &\textbf{2-node}  & \textbf{4-node} & \textbf{8-node}\\\midrule
            Spark & 0.36 M/s & 0.63 M/s & 0.94 M/s  \\
            Flink & 0.85 M/s & 1.12 M/s & 1.19 M/s  \\
        \end{tabular}
        \caption{Sustainable throughput  for windowed joins. }
        \vspace{-2em}
                 \label{tab_th_join}
    \end{table}

\begin{figure*}
	
	\centering
	\begin{subfigure}[b]{0.25\textwidth}
		\includegraphics[width=\textwidth]{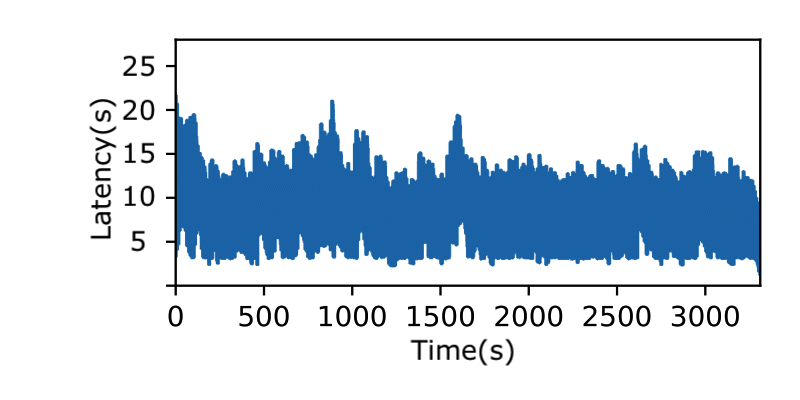}
		
		\caption{Spark, 2-node, max  throughput}
		\label{fig_spark_join_2node_th_max_ts}
	\end{subfigure}%
	\begin{subfigure}[b]{0.25\textwidth}
		\includegraphics[width=\textwidth]{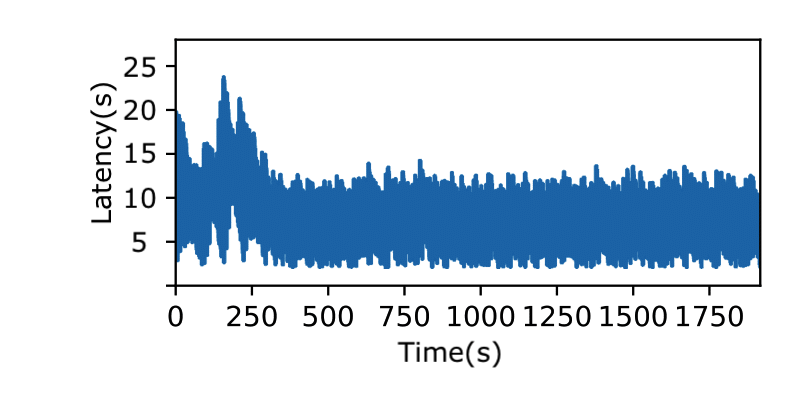}
		
		\caption{Spark, 4-node, max  throughput }
		\label{fig_spark_join_4node_th_max_ts}
	\end{subfigure}%
	\begin{subfigure}[b]{0.25\textwidth}
		\includegraphics[width=\textwidth]{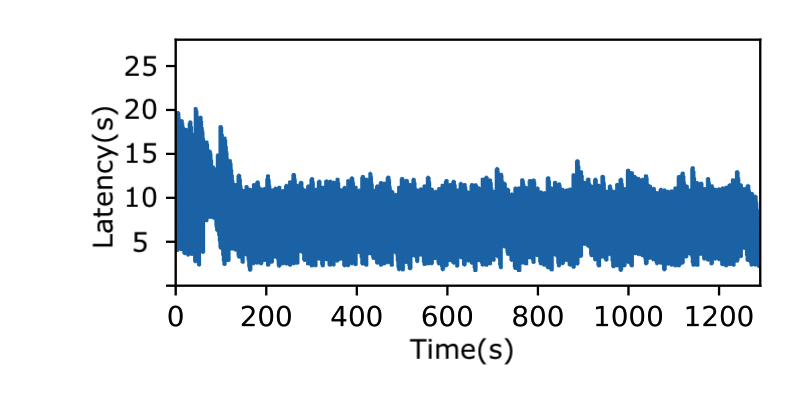}
		
		\caption{Spark, 8-node, max  throughput }
		\label{fig_spark_join_8node_th_max_ts}
	\end{subfigure}

	\begin{subfigure}[b]{0.25\textwidth}
		\includegraphics[width=\textwidth]{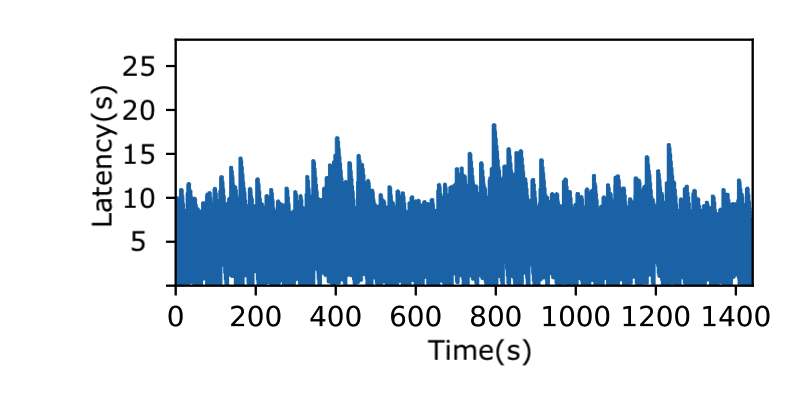}
		
		\caption{Flink, 2-node, max  throughput}
		\label{flink_join_2node_th_max_ts}
	\end{subfigure}%
	\begin{subfigure}[b]{0.25\textwidth}
		\includegraphics[width=\textwidth]{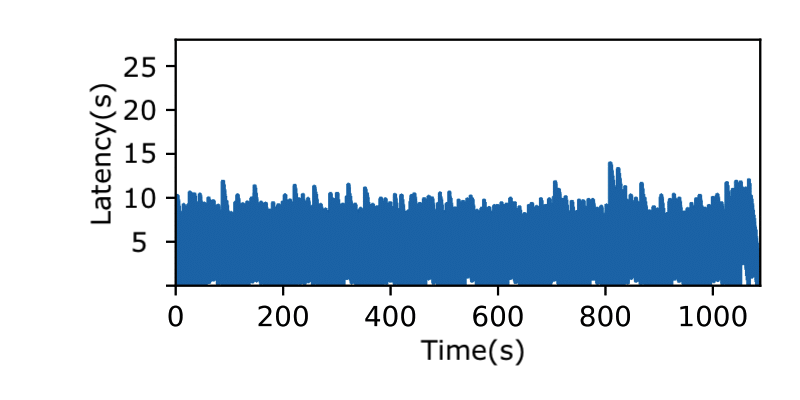}
		
		\caption{Flink, 4-node, max  throughput }
		\label{flink_join_4node_th_max_ts}
	\end{subfigure}%
	\begin{subfigure}[b]{0.25\textwidth}
		\includegraphics[width=\textwidth]{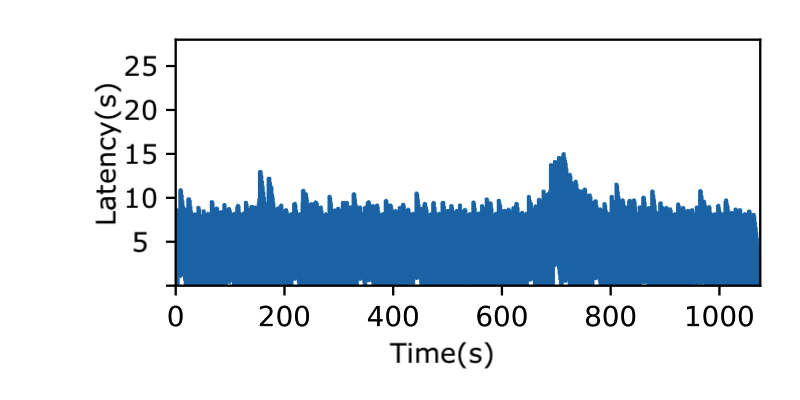}
		
		\caption{Flink, 8-node, max  throughput }
		\label{flink_join_8node_th_max_ts}
	\end{subfigure}

	\begin{subfigure}[b]{0.25\textwidth}
		\includegraphics[width=\textwidth]{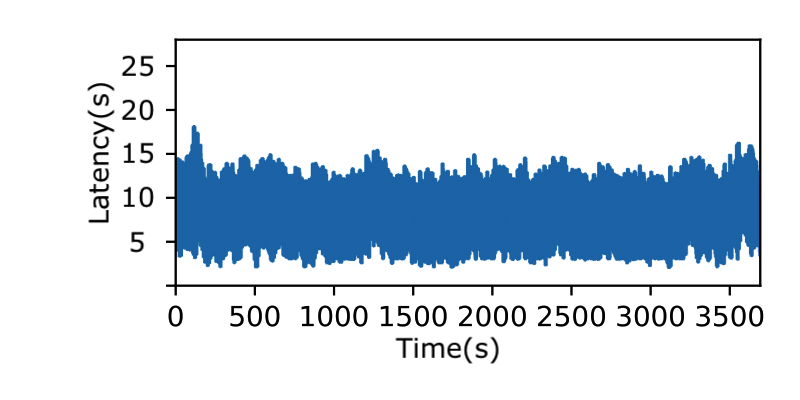}
		
		\caption{Spark, 2-node,  90\%-throughput}
	\end{subfigure}%
	\begin{subfigure}[b]{0.25\textwidth}
		\includegraphics[width=\textwidth]{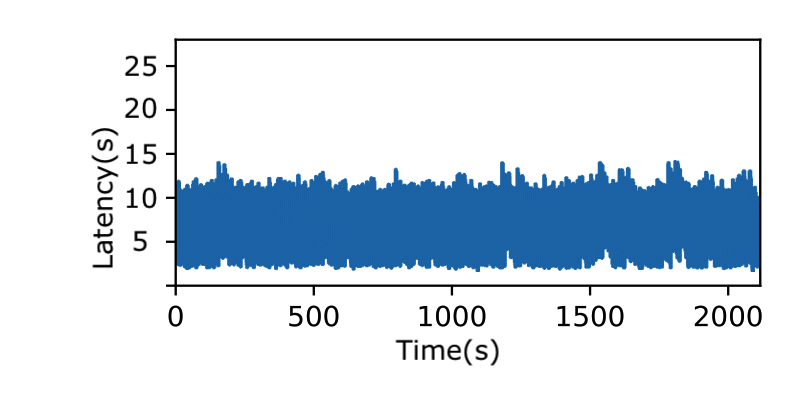}
		
		\caption{Spark, 4-node,  90\%-throughput }
	\end{subfigure}%
	\begin{subfigure}[b]{0.25\textwidth}
		\includegraphics[width=\textwidth]{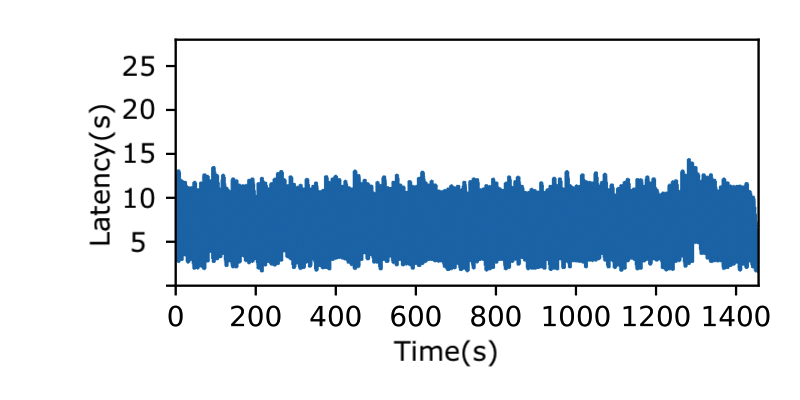}
		
		\caption{Spark, 8-node,  90\%-throughput }
	\end{subfigure}

	\begin{subfigure}[b]{0.25\textwidth}
		\includegraphics[width=\textwidth]{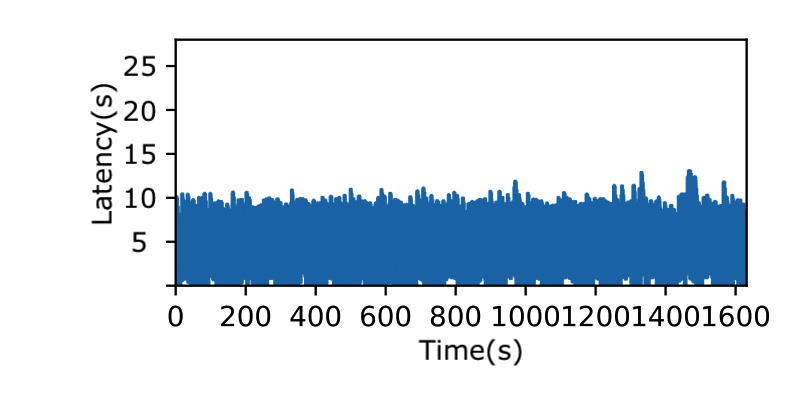}
		
		\caption{Flink, 2-node,  90\%-throughput}
		\label{flink_join_2node_th_90_ts}
	\end{subfigure}%
	\begin{subfigure}[b]{0.25\textwidth}
		\includegraphics[width=\textwidth]{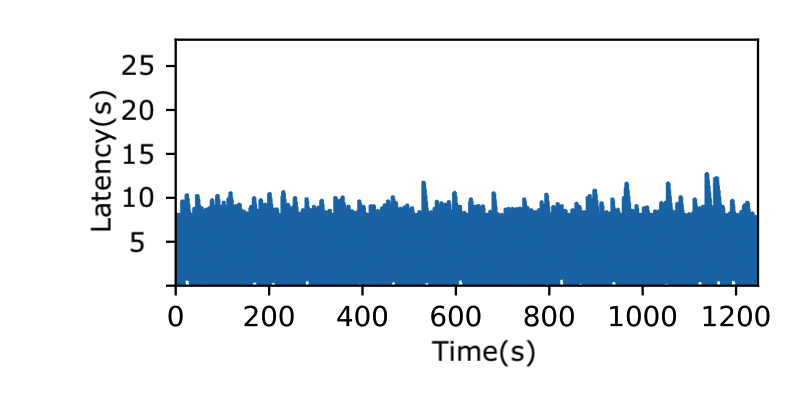}
		
		\caption{Flink, 4-node,  90\%-throughput }
	\end{subfigure}%
	\begin{subfigure}[b]{0.25\textwidth}
		\includegraphics[width=\textwidth]{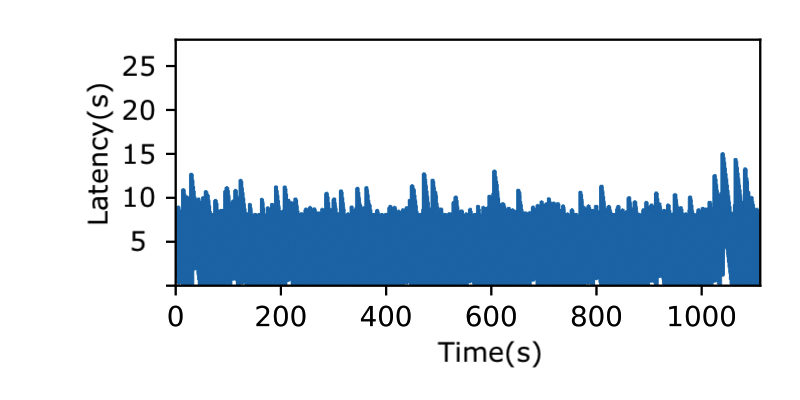}
		
		\caption{Flink, 8-node,  90\%-throughput }
	\end{subfigure}
	
	\caption{Windowed join latency distributions in time series}
	\label{fig_ts_join}
\end{figure*}

{\renewcommand{\arraystretch}{1.0}
	\begin{table*}
		\resizebox{\textwidth}{!} {\begin{tabular}{ l|c c c c | c c c c | c c c c}\toprule
				& & & \textbf{2-node}  & & & & \textbf{4-node} & & & & \textbf{8-node} \\
				& \textit{avg} & \textit{min} & \textit{max} & \textit{quantiles (90,95,99)} & \textit{avg} & \textit{min} & \textit{max} & \textit{quantiles (90,95,99)} & \textit{avg} & \textit{min} & \textit{max} & \textit{quantiles (90,95,99)}\\\midrule
				Spark & 7.7 & 1.3 & 21.6 & (11.2, 12.4, 14.7) & 6.7 & 2.1 & 23.6 & (10.2, 11.7, 15.4) & 6.2 & 1.8 & 19.9 & (9.4, 10.4, 13.2)\\
				Spark(90\%) & 7.1 & 2.1 & 17.9 & (10.3, 11.1, 12.7) & 5.8 & 1.8 & 13.9 & (8.7, 9.5, 10.7) & 5.7 & 1.7 & 14.1 & (8.6, 9.4, 10.6)\\
				Flink & 4.3 & 0.01 & 18.2 & (7.6, 8.5, 10.5) & 3.6 & 0.02 & 13.8 & (6.7, 7.5, 8.6) & 3.2 & 0.02 & 14.9 & (6.2, 7, 8.4)\\
				Flink(90\%) & 3.8 & 0.02 & 13 & (6.7, 7.5, 8.7) & 3.2 & 0.02 & 12.7 & (6.1, 6.9, 8)  & 3.2 & 0.02 & 14.9 & (6.2, 6.9, 8.3)\\
		\end{tabular}}
		\caption{Latency statistics, avg, min, max and quantiles (90, 95, 99) in seconds for windowed joins}
		\vspace{-1em}
		\label{tab_lat_join}
	\end{table*}

\para{Experiment 2: Windowed Joins.}
We use the windowed join query from Listing 1  to benchmark Spark and Flink.
Storm provides a windowing capability but there is no built-in windowed join operator. 
Initially, we tried Storm's Trident v2.0  abstrac\-tion, which has built-in windowed join features. 
However, Trident computed incorrect results as we increased the batch size. 
Moreover, there is a lack of support for Trident in the Storm community. 
As an alternative, we implemented a simple version of a windowed join in Storm. 
Comparing it with Spark and Flink, which have advanced memory and state management features, leads to unfair comparisons. 
We implemented a na\"ive join in Storm and examined the sustainable throughput to be 0.14 million events per second and measured an average latency of 2.3 seconds  on a 2-node cluster. 
However, we faced memory issues and topology stalls on larger clusters. 
As a result, we focus on Flink and Spark for the windowed join benchmarks.

First, depending on the selectivity of the input streams to the join operator, vast amount of results can be produced. Sink operators can be a bottleneck in this case. 
Second, the vast amount of results of a join operator can cause the network to be a bottleneck. 
To address this issue, we decreased the selectivity of the input streams. 
In general, the experimental results for windowed joins are similar to the experiments with windowed aggregations. 

Table \ref{tab_th_join} shows the  sustainable throughput of the systems under test. 
Flink's throughput for an 8-node cluster configuration is bounded by network bandwidth. 
The network saturation limit was 1.2 million events per second in windowed aggregations. 
The reason for the difference is that there is more network traffic as the result size is larger in windowed joins than in windowed aggregations. 
Tab\-le \ref{tab_lat_join} shows the latency statistics for windowed joins. 
We can see that  
in all cases Flink outperforms Spark in all parameters. 
To ensure the stability of the system, the runtime of each mini-batch should be less than batch size in Spark. 
Otherwise, the queued mini-batch jobs will increase over time and the system will not be able to sustain the throughput. 
However, we see from Table \ref{tab_th_join} that the latency values for Spark are higher than mini-batch duration (4 sec). 
The reason is that we are measuring the event-time latency. So, the additional latency is due to tuples' waiting in the queue.

Figure \ref{fig_ts_join} shows the windowed join latency distributions as time-series. 
In contrast to windowed aggregations, we experienced substantial fluctuations in Spark.
Also we experienced a significant latency increase in Flink when compared to windowed aggregation experiments.
The reason is that windowed joins are   more expensive than windowed aggregations. 
However, the spikes are significantly reduced with 90\% workload.

Similar to  windowed aggregations, in  windowed joins Spark's major disadvantage is having blocking operators. 
Another limitation is coordination and scheduling overhead across different RDDs. 
For example, for  windowed joins Spark  produces CoGroupedRDD, MappedValuesRDD, and FlatMappedValuesRDD in different stages of the job. 
Each of these RDDs have to wait for the parent RDDs to be complete before their initialization. 
Flink on the other hand, performs operator chaining in  query optimization part to avoid  unnecessary data migration.


\para{Experiment 3: Queries with large window.}
Window size and window slide have a significant impact on the SUT's performance. 
One interesting observation is that with  increasing window size Spark's throughput decreases significantly given the same batch size. 
For example, for the aggregation query, with window length and slide 60 seconds, Spark's throughput decreases by 2 times and $avg$ latency increases by 10 times given a 4-seconds batch size. 
However, with higher batch sizes, Spark can handle  higher workloads and big windows sacrificing low latencies. 
We find that the main reason for Spark's decreasing performance is caching. 
We cache the windowed operation results to be usable in later stages of a job. 
However, especially with windowed join queries, the cache operation consumes the memory aggressively. 
Internally when a task gets a record for processing, it checks if the record is marked for caching. 
If yes, all the following records of the particular RDD will be sent to memory store of block manager. 
Spark will spill the memory store to disk once it is full. 
As a result, we disabled the caching. 
However, then we experienced the performance decreased due to the repeated  computation. 
Finally, after implementing Inverse Reduce Function, to account for old data going out of window, we managed to overcome this performance issues.

Storm, on the other hand, can handle the large window  operations if the user has advanced data structures that can spill to disk when needed. 
Otherwise, we encountered memory exceptions. 
Flink (as well as Spark) has built-in data structures that can spill to disk when needed. 
However, this does not apply for the operations inside the UDFs, as Flink treats UDFs as blackbox. 
Analyzing the implementation of windowed aggregates in Flink we conclude that it cannot share aggregate results among different sliding windows. 
As Flink computes aggregates on-the-fly and not after window closes, this limitation is not a dominating factor in overall performance.

\para{Experiment 4: Data skew.}
Data skew is yet another concern to be addressed  by SDPSs.
We analyzed the systems under test with extreme skew, namely their ability to handle data of a single key. 
In Flink and Storm, the performance of the system is bounded by the performance of a single slot of a machine, meaning it does not scale. The reason lies under the design option of a system. For aggregation query, we measured the throughput 0.48 M/s for Flink and 0.2 M/s for Storm. As we mentioned above, these measurements does not improve when SUT scales.

Spark, on the other hand, can handle skewed data efficiently. We experienced 0.53 M/s sustainable throughput for Spark in a 4-node cluster. 
Spark outperforms both engines  for the  aggregation query  in a 4- or more node cluster. 
For the join query, on the other hand, both Spark and Flink cannot  handle skewed data well. 
That is, Flink often becomes unresponsive in this test. 
Spark, on the other hand, exhibits very high latencies. 
The main reason is that the memory is consumed quite fast and backpressure mechanism lacks to perform efficiently. 
In these experiments, we decreased the join operator's selectivity to make sure that data transfer and I/O are not a bottleneck.

One reason for performance difference between Spark and Flink with skewed data  lies on how the systems compute aggregations. 
Flink and Storm use one slot per operator instance. So, if the input data is skewed, this architecture can cause performance issues. 
Spark  has a slightly different architecture.
In Spark, forcing all partitions  to send their reduced value to a specific computing slot can easily cause a network to become a bottleneck when partition size is big. Therefore, Spark adopts tree reduce and tree aggregate communication pattern to minimize the communication and data shuffling. This is the main reason that makes Spark perform better with skewed input data.

\begin{figure*}
	\centering
	
	\begin{subfigure}[b]{0.19\textwidth}
		\includegraphics[width=\textwidth]{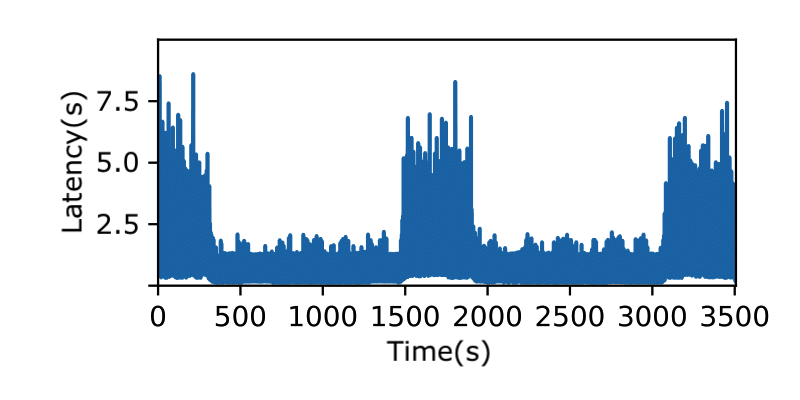}

		\caption{Storm aggregation }
	\end{subfigure}\hspace{-1mm}%
	\begin{subfigure}[b]{0.19\textwidth}
		\includegraphics[width=\textwidth]{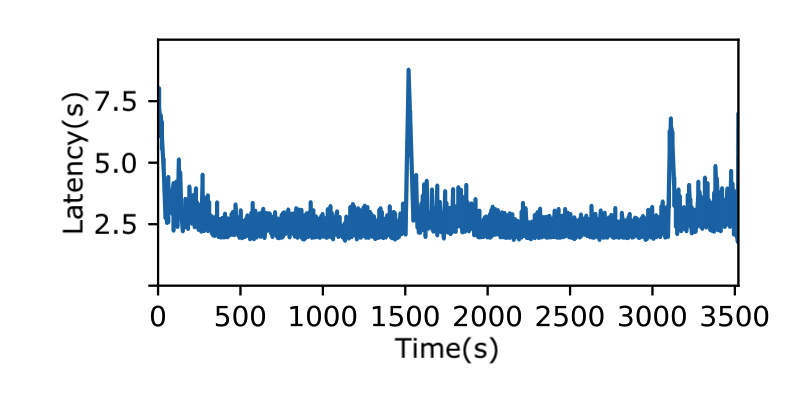}

		\caption{ Spark aggregation} 
	\end{subfigure}\hspace{-1mm}%
	\begin{subfigure}[b]{0.19\textwidth}
		\includegraphics[width=\textwidth]{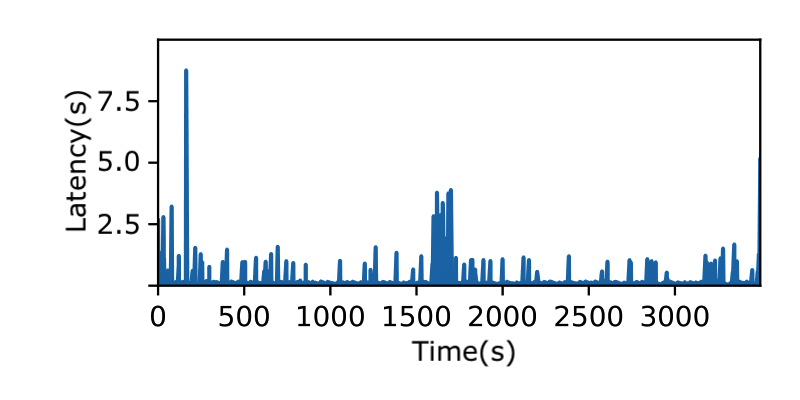}

		\caption{ Flink aggregation}
	\end{subfigure}\hspace{-1mm}%
	\begin{subfigure}[b]{0.19\textwidth}
		
		\includegraphics[width=\textwidth]{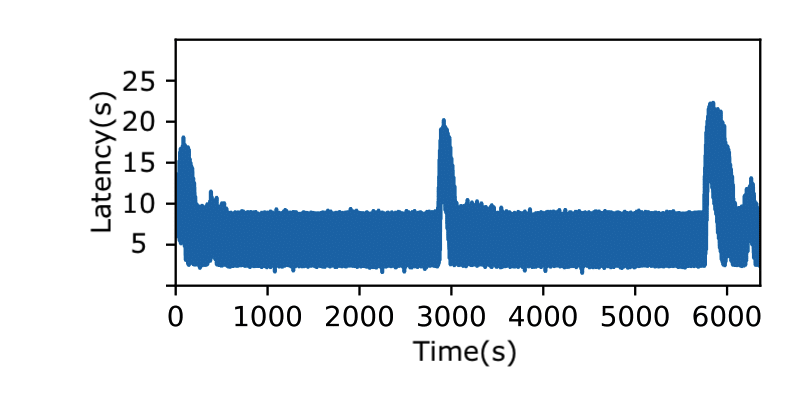}

		\caption{Spark join}
		\label{peaks_spark_join}
	\end{subfigure}\hspace{-1mm}%
	\begin{subfigure}[b]{0.19\textwidth}
		
		\includegraphics[width=\textwidth]{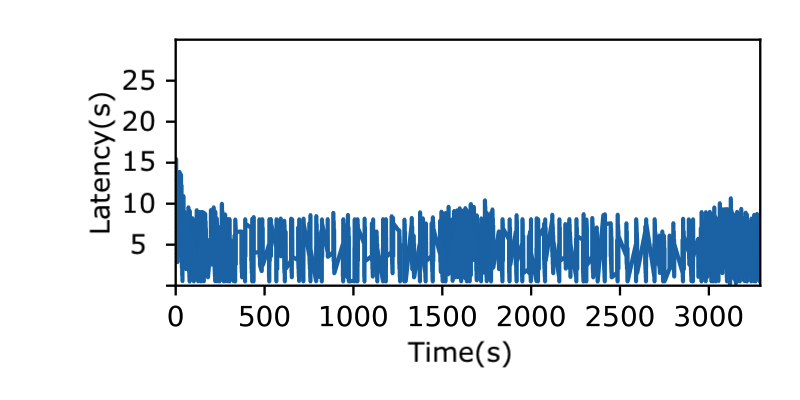}
		
		\caption{ Flink join} 
	\end{subfigure}
	\caption{Event-time latency on workloads with fluctuating data arrival rate.}
	\label{peaks}
\end{figure*}

\begin{figure}
	\centering
	
	\begin{subfigure}[b]{0.24\textwidth}
		\includegraphics[width=\textwidth]{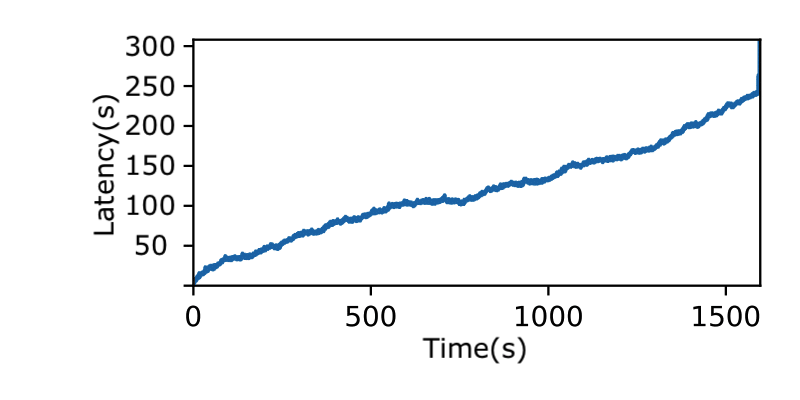}
		
		\caption{Event time}
	\end{subfigure}\hspace{-1mm}%
	\begin{subfigure}[b]{0.24\textwidth}
		\includegraphics[width=\textwidth]{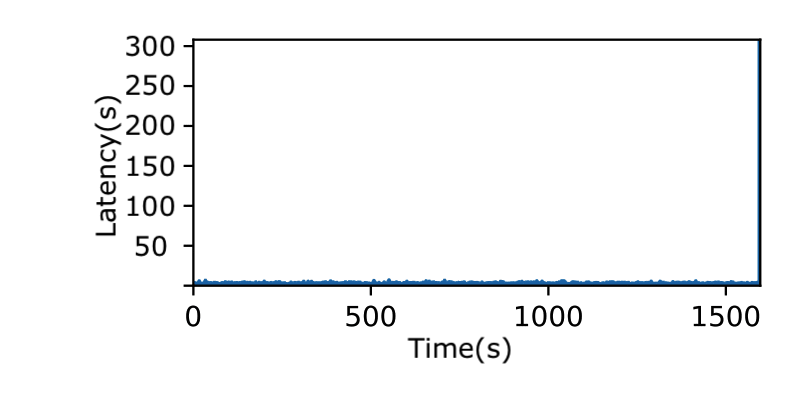}
		
		\caption{Processing time}
	\end{subfigure}
	
	\caption{Comparison between event  and processing-time  latency of Spark with unsustainable throughput}

	\label{more_th}
\end{figure}

\begin{figure*}
	\centering
	
	\begin{subfigure}[b]{0.28\textwidth}
		\includegraphics[width=\textwidth]{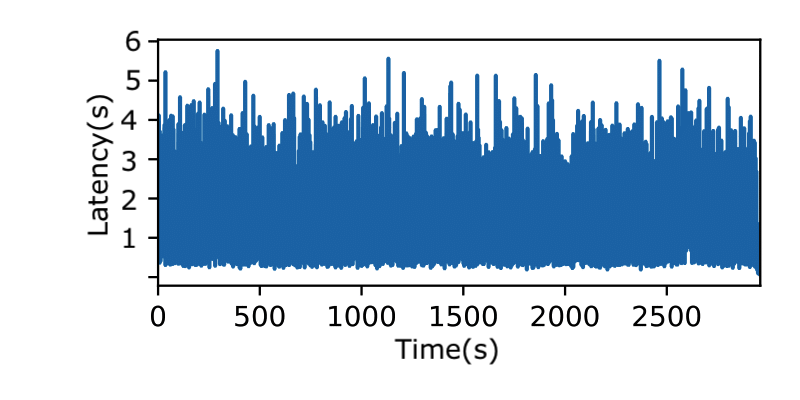}
		\includegraphics[width=\textwidth]{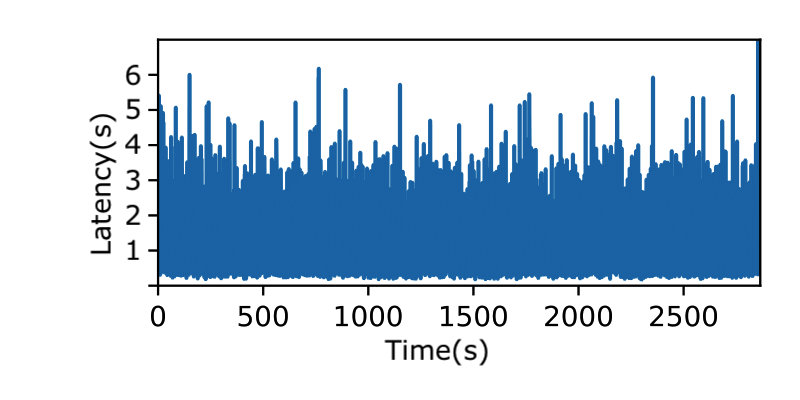}
		
		\caption{Storm} 
	\end{subfigure}\hspace{-1mm}%
	\begin{subfigure}[b]{0.28\textwidth}
		\includegraphics[width=\textwidth]{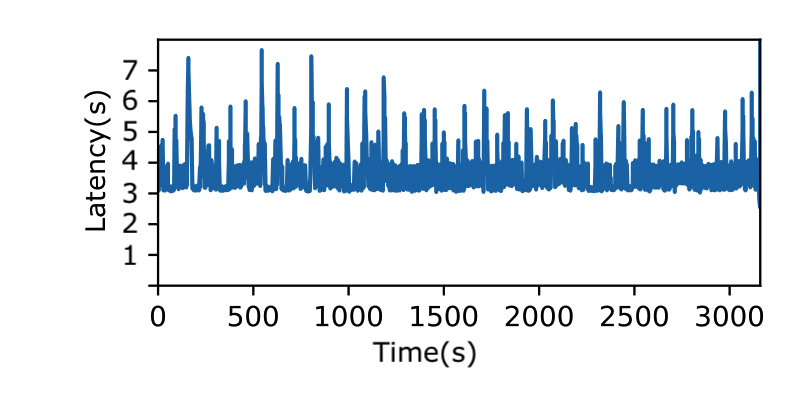}
		\includegraphics[width=\textwidth]{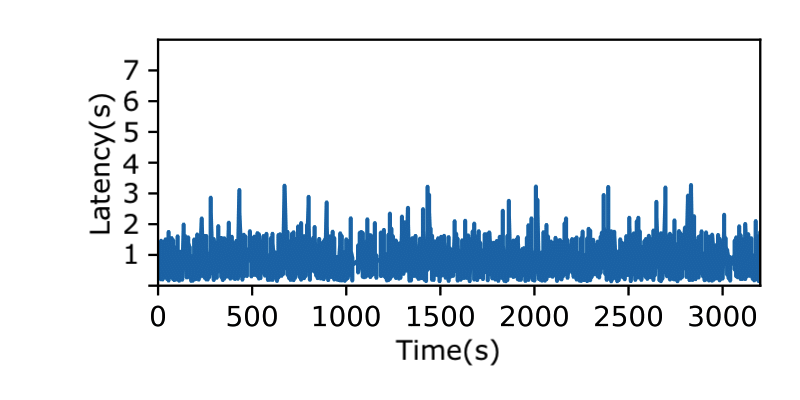}
		
		\caption{Spark} 
	\end{subfigure}\hspace{-1mm}%
	\begin{subfigure}[b]{0.28\textwidth}
		\includegraphics[width=\textwidth]{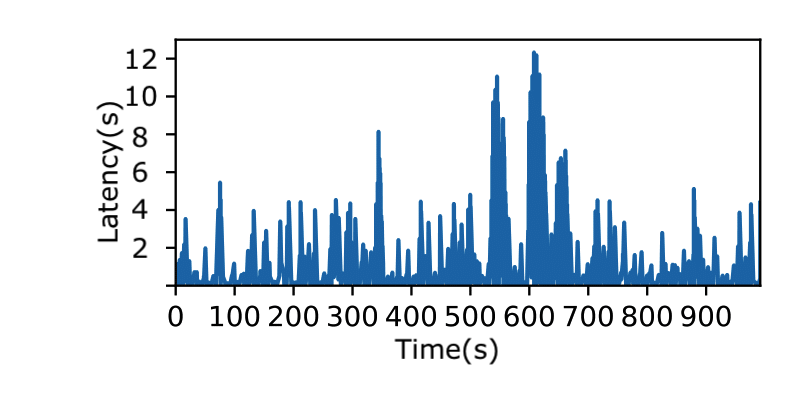}
		\includegraphics[width=\textwidth]{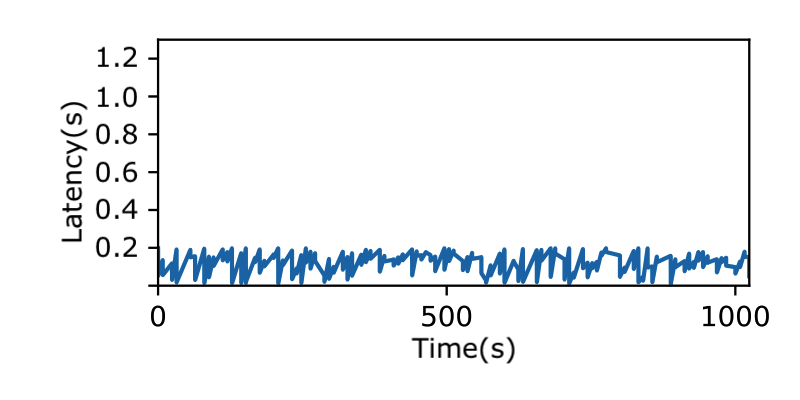}
		
		\caption{Flink}
	\end{subfigure}\hspace{-1mm}%
	\caption{Comparison between event (top row) and processing-time (bottom row) latency}
	\label{event_vs_system_latency}
\end{figure*}

\begin{figure*}
	\centering
	
	\begin{subfigure}[b]{0.3\textwidth}
		\includegraphics[width=\textwidth]{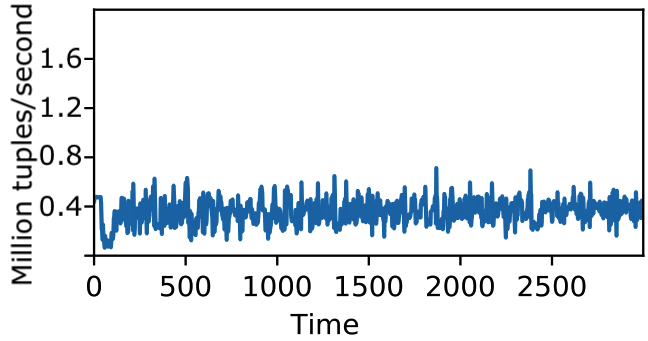}

		\caption{Storm} 
	\end{subfigure}%
	\begin{subfigure}[b]{0.3\textwidth}
		\includegraphics[width=\textwidth]{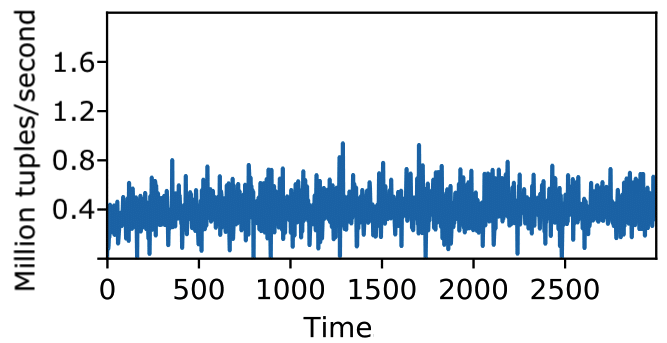}

		\caption{Spark} 
	\end{subfigure}%
	\begin{subfigure}[b]{0.3\textwidth}
		\includegraphics[width=\textwidth]{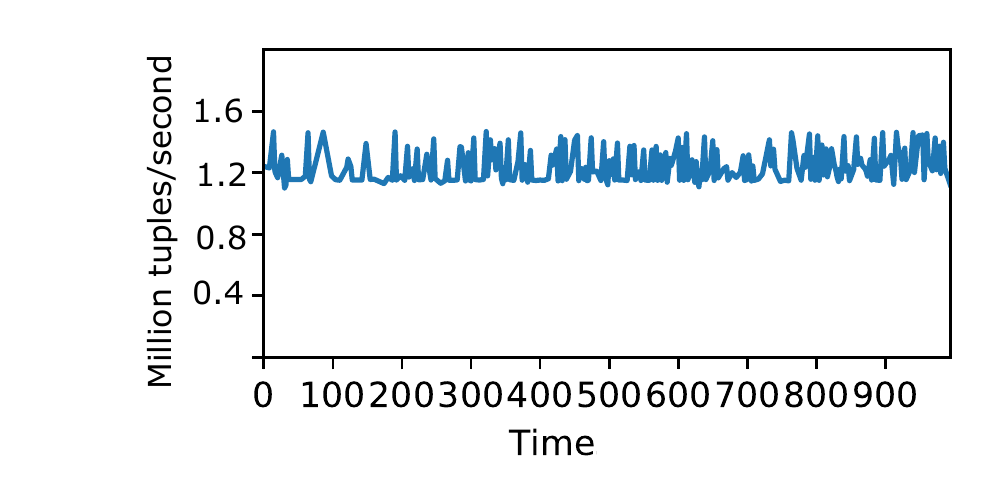}

		\caption{ Flink} 
	\end{subfigure}
	
	\caption{Throughput graphs of systems under test}
	\label{pull_graph}
\end{figure*}

Spark performs worse with less computing units (3- or less node) than Flink or Storm because it has blocking operators. As the system scales out, this limitation becomes no longer a dominating factor. Flink avoids blocking operators. For example, the reduce operation is a non-blocking operator in Flink. As a result, the system sacrifices some use-cases which need blocking reduce, to achieve a better performance. Internally, Storm also has a similar architecture; however, the semantics of an operator is highly dependent on its implementation. For example, one implementation of window reduce operator can output the results continuously, while another can chose to perform so in bulk. 

\begin{figure*}
	\centering
	
	\begin{subfigure}[b]{0.35\textwidth}
		\includegraphics[width=\textwidth]{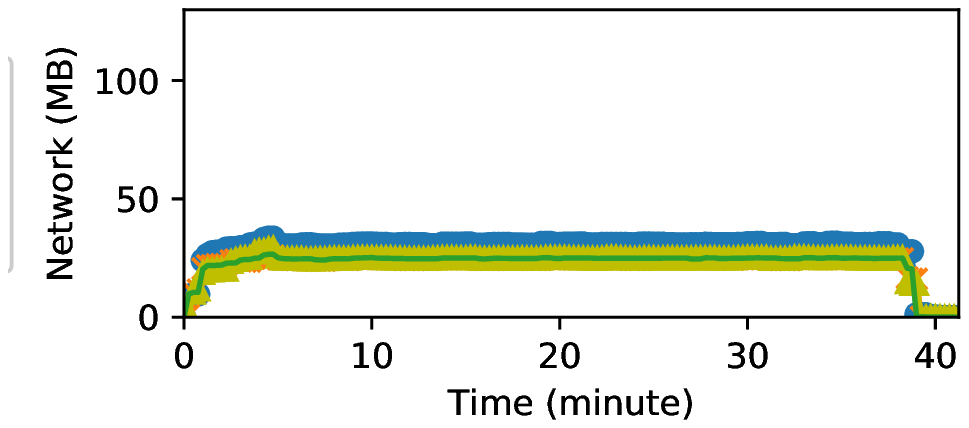}
		\includegraphics[width=\textwidth]{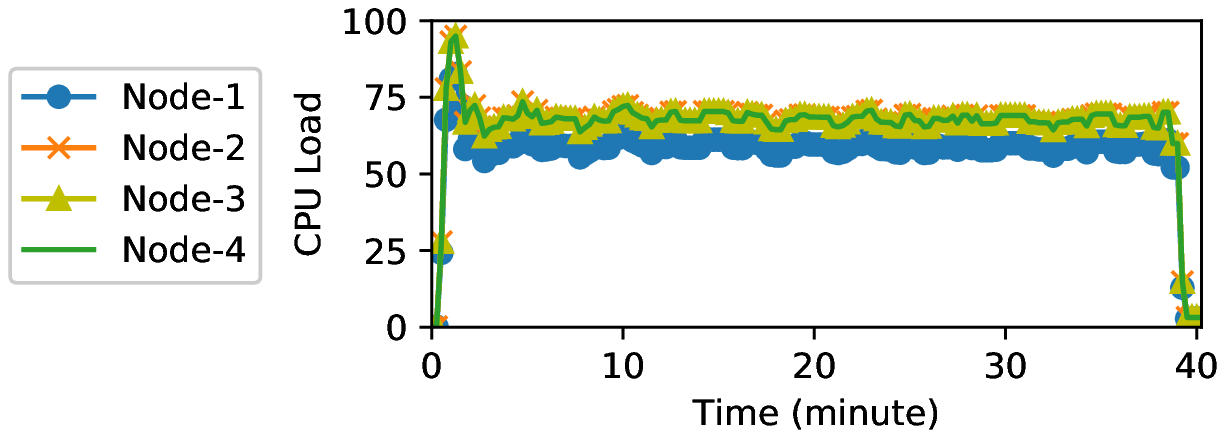}
		\caption{Storm }
	\end{subfigure}%
	\begin{subfigure}[b]{0.27\textwidth}
		\includegraphics[width=\textwidth]{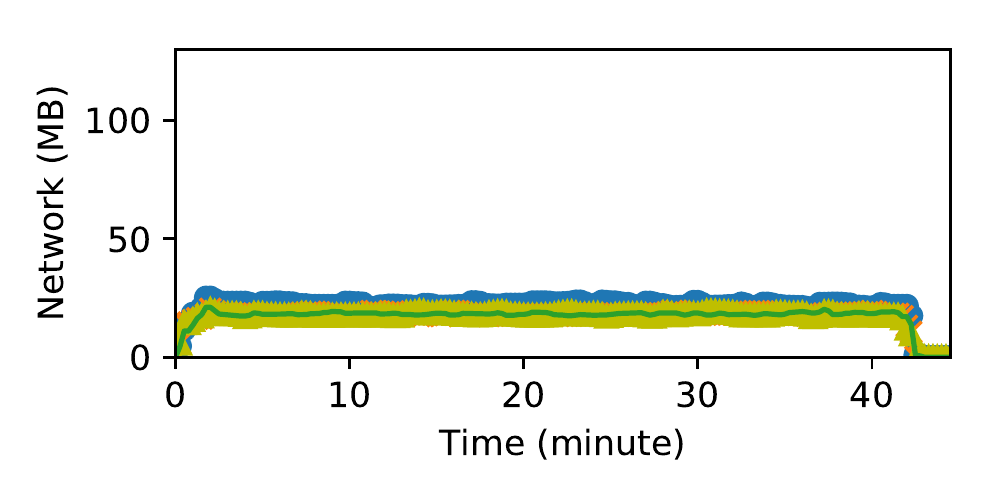}
		\includegraphics[width=\textwidth]{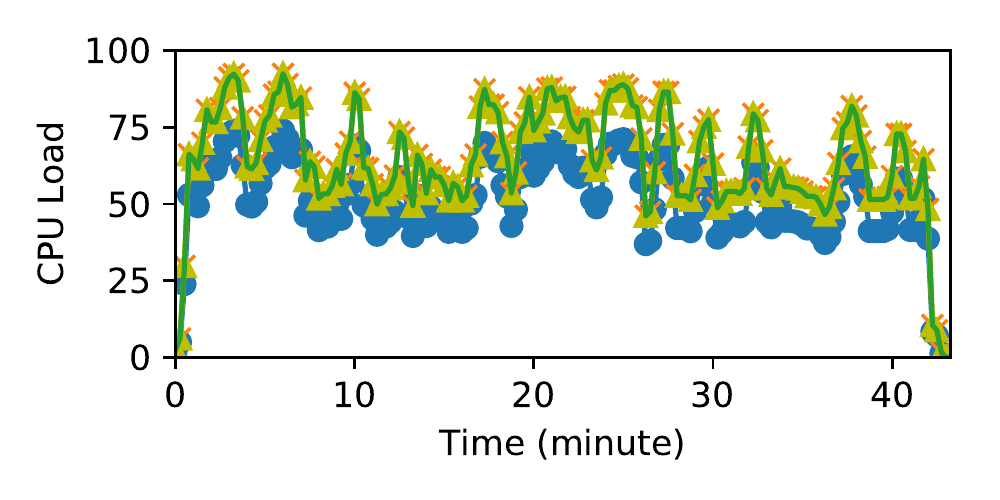}
		\caption{Spark }
	\end{subfigure}%
	\begin{subfigure}[b]{0.27\textwidth}
		\includegraphics[width=\textwidth]{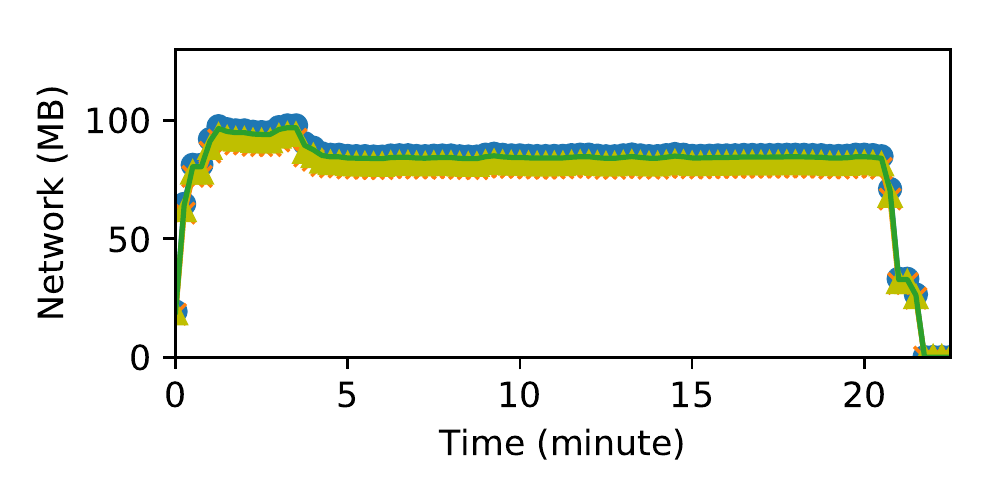}
		\includegraphics[width=\textwidth]{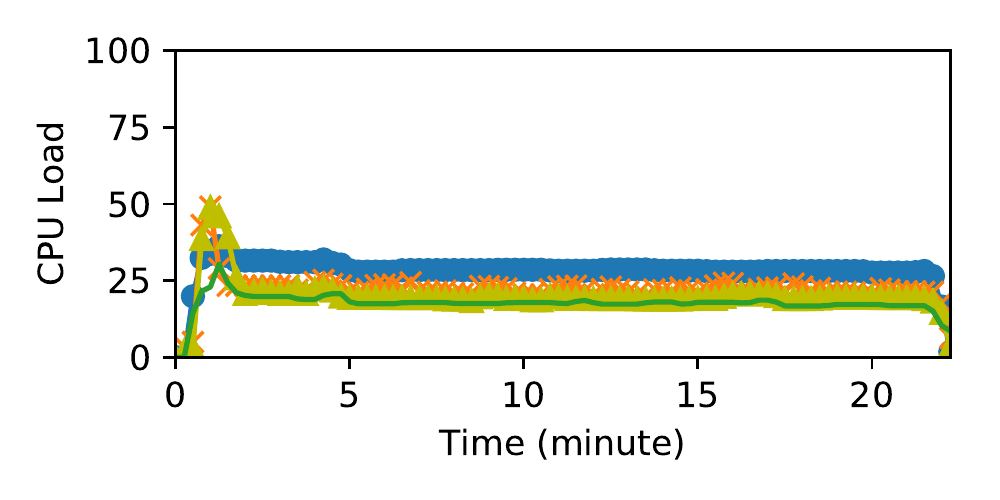}
		\caption{Flink }
	\end{subfigure}
	
	\caption{Network (top row) CPU  (bottom row) usages  of systems in a 4-node cluster.}
	\vspace{-2em}
	\label{fig_cpu_network_metrics}
\end{figure*}

\begin{figure}
	\centering
		\begin{subfigure}[b]{0.34\textwidth}
				\includegraphics[width=\textwidth]{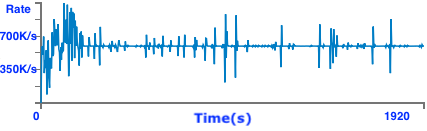}

		\caption{Throughput}
	\end{subfigure}
	\begin{subfigure}[b]{0.34\textwidth}
				\includegraphics[width=\textwidth]{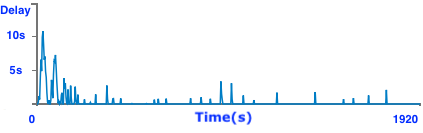}

		\caption{Scheduler delay} 
	\end{subfigure}\hspace{-1mm}%
	
	\caption{Scheduler delay (top row) vs. throughput (bottom row) in Spark.}
	\label{scheduler}
\end{figure}

\para{Experiment 5: Fluctuating workloads.}
We analyze the systems under test with fluctuating workloads. 
We simulate spikes for all systems both for aggregation and join queries. 
We start the benchmark with a workload of 0.84 M/s  then decrease it to 0.28 M/s and increase again after a while. 
As we can see from Figure \ref{peaks}, Storm is  most susceptible system for fluctuating workloads. 
Spark and Flink have competitive behavior with windowed aggregations. 
However, for windowed joins, Flink can handle spikes better. 
One reason behind this behavior is the difference between the systems' backpressure mechanism. 
As we mentioned above, Spark can be thought as a chain of jobs with multiple stages. 
Starting from the final RDD, each stage checks its backwards to ensure lazy evaluation. 
As a result, once the stage is overloaded passing this information to upstream stages works in the order of job stage execution time; however, this time is in the order of tuples in Flink. 
We conduct experiments with different cluster and buffer sizes as well. As we increase the buffer or cluster size  the spikes  get smoother; however, the overall $avg$ latency increases.

\para{Experiment 6: Event-time vs processing-time latency.}
\label{event_vs_system}
Figure \ref{event_vs_system_latency} shows the comparison between the processing-time and event-time latency. 
We conduct  experiments with aggregation query (8s,4s) on a 2-node cluster. 
Even with a small cluster size,  we can see from Figure \ref{event_vs_system_latency} that there is a significant difference between event and  processing-time  latencies. 
As a result, we can see that with Spark, input tuples spend most of the time in driver queues. 
We didn't examine any significant changes in results with different cluster configurations and with join query.

To emphasize the necessity of our definition of latency, we draw reader's attention to Figure \ref{more_th}, which shows  event time and processing-time latencies for Spark when the system is extremely overloaded. 
This is not a specific behavior for Spark but we observed similar graphs for all systems. 
As we can see from the figures, the processing-time latency is significantly lower than event-time latency. 
The reason is that when the SUT gets overloaded, it starts backpressure and lowers the data ingestion rate to stabilize  the end-to-end system latency. 
We can see that the SUT accomplished this goal as the latency stays stable. 
However, the event-time latency keeps increasing as the input tuples wait  in the queues. 
This is just one scenario where we can draw \textit{unrealistic} or \textit{incorrect} conclusions when using traditional processing-time latency for streaming systems.

	
		
		
	


\para{Experiment 7: Observing backpressure.}
Backpressure is shown in Figures  \ref{fig_spark_join_2node_th_max_ts}, \ref{fig_spark_join_4node_th_max_ts}, \ref{fig_spark_join_8node_th_max_ts}, 	\ref{flink_join_2node_th_max_ts}, \ref{flink_join_4node_th_max_ts}, \ref{flink_join_8node_th_max_ts}, and \ref{flink_agg_2node_th_max_ts}. 
Moreover, our driver can also observe short-term spikes (Figures \ref{fig_storm_agg_8node_th_max_ts}, \ref{peaks_spark_join}) and  continuous fluctuations (Figure \ref{flink_join_2node_th_90_ts}). 
Furthermore, we can observe a wide range of sustainable $avg$ latencies from 0.2 to 6.2 seconds and from 0.003 seconds $min$ latency to 19.9 seconds $max$ latency.


\para{Experiment 8: Throughput graphs.}
While analyzing the  performance of systems it is essential to inspect their throughput over time. 
As we separate the throughput calculation clearly from the SUT, we retrieve this metric from the driver. 
Figure \ref{pull_graph} shows the sustainable throughput graphs for the aggregation query (8s, 4s). 
We examined  similar behavior in other window settings and for the  join query as long as the workload is maximum sustainable. 
As we can see, Spark and Storm pulls data from the source in a more fluctuating way  than Flink. 
Despite having a high data pull rate or throughput, Flink has less fluctuations. 
When we lower the workload, both Flink and Spark have stable data pull rates; however, Storm still exhibits significant fluctuations. 

The reason for highly fluctuating throughput for Storm is that it lacks an efficient backpressure mechanism to find a near-constant data ingestion rate. 
The main reason for fluctuation in Spark is the deployment of several jobs at the same batch interval. 
Spark has an $action()$ method on each RDD; the number of jobs at any time will be equal to the number of $action()$ method calls. 
Each job retrieves the data into its input buffers and fires. 
Until a job is finished, its input rate is limited. 
It is the DagScheduler's job to coordinate and schedule all running jobs. 
As a result, we can see a highly fluctuating throughput for Spark. 
Flink, on the other hand, benefits from its internally incremental computation mechanism (like Spark), tuple at a time semantics and efficient backpressure mechanism.

\para{Resource usage statistics.} Figure \ref{fig_cpu_network_metrics} shows the resource usages of the SUTs. 
The below graphs in  Figure \ref{fig_cpu_network_metrics} show the CPU load during the experiment. 
The above graphs in Figure \ref{fig_cpu_network_metrics}  show the network usage of the SUTs. 
Because the overall result is similar, we show the systems' resource utilization graphs  for the aggregation query  in a 4-node cluster. 
Because Flink's performance is bounded by the network, we can see that CPU load is least. 
Storm and Spark, on the other hand, use approximately 50\% more CPU clock cycles than Flink. As we can see from Figure \ref{scheduler}, the scheduler overhead is one bottleneck for Spark's performance. 
Initially, Spark ingests more tuples  than it can sustain. 
Because of the scheduler delay, backpressure fires and limits the input rate. 
Whenever there is even a short spike in the input rate, we can observe a similar behavior in the scheduler delay.

\subsection{Discussion}

If a stream contains skewed data, then Spark is the best choice (Experiment 4).  Both Flink and Spark are very robust to fluctuations in the data arrival rate in aggregation workloads (Experiment 5).  For fluctuations in the data arrival rate on join queries, on the other hand, Flink behaves better (Experiment 5). In general, if the average latency is a priority, then Flink is the best choice (Experiments 1 and 2). On the other hand, even with higher average latency, Spark manages to bound latency better than others (Experiments 1 and 2). If a use-case contains large windows, Flink can have higher throughput with a low latency (Experiment 3). Overall, we observe that Flink has a better overall throughput both for aggregation and join queries. 
We give a definition of event and processing time latency and show the significant difference them  (Experiment 6).

\subsection{Future work}

We plan to extend our framework along the lines of TPC database benchmarks. The main intuition is  to define both a workload of queries that should be concurrently executed  and then base the benchmark on a small number of operators that are part of that workload. In addition, we are developing a generic interface that users can plug into any stream data processing system, such as Apache Samza, Heron, and Apache Apex,  in order to facilitate and simplify benchmark SDPSs. Moreover, in depth analysis of trading SUT's increased functionality, like exactly once processing or out-of-order and late arriving data management, over better throughput/latency is another open challenge to explore.

\section{Conclusions}
\label{conc}

Responding to an increasing need for real-time data processing in industry, we build a novel framework for benchmarking streaming engines with online video game scenarios. 
We identify current challenges in this area and build our benchmark to evaluate them. 
First, we give the  definition of latency of a stateful operator and a methodology to measure it. 
The solution is lightweight and does not require the use of additional systems. 
Second, we completely separate the systems under test from the driver. 
Third, we introduce a simple and novel technique to conduct  experiments with the highest sustainable workloads.
We conduct extensive experiments with the three major distributed, open source stream processing engines - Apache Storm, Apache Spark, and Apache Flink. 
In the experiments, we can see that each system has specific advantages and challenges. 
We provide a set of rules in our discussion part that can be used as a guideline to determine the requirements for a use-case.

\section*{Acknowledgments}
This work has been supported by the European Commission through Proteus (ref. 687691) and Streamline (ref. 688191) and by the German Ministry for Education and Research as Berlin Big Data Center BBDC (funding mark 01IS14013A).

\bibliographystyle{IEEEtran}
\bibliography{icde2018}

\begin{thebibliography}{10}
\providecommand{\url}[1]{#1}
\csname url@samestyle\endcsname
\providecommand{\newblock}{\relax}
\providecommand{\bibinfo}[2]{#2}
\providecommand{\BIBentrySTDinterwordspacing}{\spaceskip=0pt\relax}
\providecommand{\BIBentryALTinterwordstretchfactor}{4}
\providecommand{\BIBentryALTinterwordspacing}{\spaceskip=\fontdimen2\font plus
\BIBentryALTinterwordstretchfactor\fontdimen3\font minus
  \fontdimen4\font\relax}
\providecommand{\BIBforeignlanguage}[2]{{%
\expandafter\ifx\csname l@#1\endcsname\relax
\typeout{** WARNING: IEEEtran.bst: No hyphenation pattern has been}%
\typeout{** loaded for the language `#1'. Using the pattern for}%
\typeout{** the default language instead.}%
\else
\language=\csname l@#1\endcsname
\fi
#2}}
\providecommand{\BIBdecl}{\relax}
\BIBdecl

\bibitem{toshniwal2014storm}
A.~Toshniwal, S.~Taneja, A.~Shukla, K.~Ramasamy, J.~M. Patel, S.~Kulkarni,
  J.~Jackson, K.~Gade, M.~Fu, J.~Donham \emph{et~al.}, ``Storm@ twitter,'' in
  \emph{Proceedings of the 2014 ACM SIGMOD}.\hskip 1em plus 0.5em minus
  0.4em\relax ACM, 2014, pp. 147--156.

\bibitem{zaharia2012discretized}
M.~Zaharia, T.~Das, H.~Li, S.~Shenker, and I.~Stoica, ``Discretized streams: an
  efficient and fault-tolerant model for stream processing on large clusters,''
  in \emph{Presented as part of the}, 2012.

\bibitem{carbone2015apache}
P.~Carbone, S.~Ewen, S.~Haridi, A.~Katsifodimos, V.~Markl, and K.~Tzoumas,
  ``Apache flink: Stream and batch processing in a single engine,'' \emph{IEEE
  Data Engineering Bulletin}, 2015.

\bibitem{huang2011hibench}
S.~Huang, J.~Huang, J.~Dai, T.~Xie, and B.~Huang, ``The hibench benchmark
  suite: Characterization of the mapreduce-based data analysis,'' in \emph{New
  Frontiers in Information and Software as Services}.\hskip 1em plus 0.5em
  minus 0.4em\relax Springer, 2011, pp. 209--228.

\bibitem{white2012hadoop}
T.~White, \emph{Hadoop: The definitive guide}.\hskip 1em plus 0.5em minus
  0.4em\relax " O'Reilly Media, Inc.", 2012.

\bibitem{li2015sparkbench}
M.~Li, J.~Tan, Y.~Wang, L.~Zhang, and V.~Salapura, ``Sparkbench: a
  comprehensive benchmarking suite for in memory data analytic platform
  spark,'' in \emph{Proceedings of the 12th ACM International Conference on
  Computing Frontiers}.\hskip 1em plus 0.5em minus 0.4em\relax ACM, 2015.

\bibitem{ghazal2013bigbench}
A.~Ghazal, T.~Rabl, M.~Hu, F.~Raab, M.~Poess, A.~Crolotte, and H.-A. Jacobsen,
  ``Bigbench: towards an industry standard benchmark for big data analytics,''
  in \emph{Proceedings of the 2013 ACM SIGMOD}.\hskip 1em plus 0.5em minus
  0.4em\relax ACM, 2013, pp. 1197--1208.

\bibitem{wang2014bigdatabench}
L.~Wang, J.~Zhan, C.~Luo, Y.~Zhu, Q.~Yang, Y.~He, W.~Gao, Z.~Jia, Y.~Shi,
  S.~Zhang \emph{et~al.}, ``Bigdatabench: A big data benchmark suite from
  internet services,'' in \emph{2014 IEEE HPCA}.\hskip 1em plus 0.5em minus
  0.4em\relax IEEE, 2014, pp. 488--499.

\bibitem{marcu2016spark}
O.-C. Marcu, A.~Costan, G.~Antoniu, and M.~S. P{\'e}rez, ``Spark versus flink:
  Understanding performance in big data analytics frameworks,'' in
  \emph{Cluster 2016-The IEEE 2016 International Conference on Cluster
  Computing}, 2016.

\bibitem{chintapalli2016benchmarking}
S.~Chintapalli, D.~Dagit, B.~Evans, R.~Farivar, T.~Graves, M.~Holderbaugh,
  Z.~Liu, K.~Nusbaum, K.~Patil, B.~J. Peng \emph{et~al.}, ``Benchmarking
  streaming computation engines: Storm, flink and spark streaming,'' in
  \emph{IEEE International Parallel and Distributed Processing Symposium
  Workshops}.\hskip 1em plus 0.5em minus 0.4em\relax IEEE, 2016, pp.
  1789--1792.

\bibitem{kreps2011kafka}
J.~Kreps, N.~Narkhede, J.~Rao \emph{et~al.}, ``Kafka: A distributed messaging
  system for log processing,'' in \emph{Proceedings of the NetDB}, 2011, pp.
  1--7.

\bibitem{carlson2013redis}
J.~L. Carlson, \emph{Redis in Action}.\hskip 1em plus 0.5em minus 0.4em\relax
  Manning Publications Co., 2013.

\bibitem{karamel}
``{Karamel, Orchestrating Chef Solo},'' \url{http://storm.apache.org},
  accessed: 2017-01-28.

\bibitem{perera2016reproducible}
S.~Perera, A.~Perera, and K.~Hakimzadeh, ``Reproducible experiments for
  comparing apache flink and apache spark on public clouds,'' \emph{arXiv
  preprint arXiv:1610.04493}, 2016.

\bibitem{dataartisans}
DataArtisans, ``{Extending the Yahoo! Streaming Benchmark},''
  \url{http://data-artisans.com/extending-the-yahoo-streaming-benchmark/},
  2016, [Online; accessed 19-Nov-2016].

\bibitem{lopez2016performance}
M.~A. Lopez, A.~Lobato, and O.~Duarte, ``A performance comparison of
  open-source stream processing platforms,'' in \emph{IEEE Globecom}, 2016.

\bibitem{shukla2016benchmarking}
A.~Shukla and Y.~Simmhan, ``Benchmarking distributed stream processing
  platforms for iot applications,'' \emph{arXiv preprint arXiv:1606.07621},
  2016.

\bibitem{arasu2004linear}
A.~Arasu, M.~Cherniack, E.~Galvez, D.~Maier, A.~S. Maskey, E.~Ryvkina,
  M.~Stonebraker, and R.~Tibbetts, ``Linear road: a stream data management
  benchmark,'' in \emph{Proceedings of the VLDB-Volume 30}.\hskip 1em plus
  0.5em minus 0.4em\relax VLDB Endowment, 2004, pp. 480--491.

\bibitem{lu2014stream}
R.~Lu, G.~Wu, B.~Xie, and J.~Hu, ``Streambench: Towards benchmarking modern
  distributed stream computing frameworks,'' in \emph{IEEE/ACM UCC}.\hskip 1em
  plus 0.5em minus 0.4em\relax IEEE, 2014.

\bibitem{neumeyer2010s4}
L.~Neumeyer, B.~Robbins, A.~Nair, and A.~Kesari, ``S4: Distributed stream
  computing platform,'' in \emph{2010 IEEE International Conference on Data
  Mining Workshops}.\hskip 1em plus 0.5em minus 0.4em\relax IEEE, 2010, pp.
  170--177.

\bibitem{qian2013timestream}
Z.~Qian, Y.~He, C.~Su, Z.~Wu, H.~Zhu, T.~Zhang, L.~Zhou, Y.~Yu, and Z.~Zhang,
  ``Timestream: Reliable stream computation in the cloud,'' in
  \emph{Proceedings of the 8th ACM European Conference on Computer
  Systems}.\hskip 1em plus 0.5em minus 0.4em\relax ACM, 2013, pp. 1--14.

\bibitem{ranjan2014streaming}
R.~Ranjan, ``Streaming big data processing in datacenter clouds,'' \emph{IEEE
  Cloud Computing}, vol.~1, no.~1, pp. 78--83, 2014.

\bibitem{akidau2015dataflow}
T.~Akidau, R.~Bradshaw, C.~Chambers, S.~Chernyak, R.~J.
  Fern{\'a}ndez-Moctezuma, R.~Lax, S.~McVeety, D.~Mills, F.~Perry, E.~Schmidt
  \emph{et~al.}, ``The dataflow model: a practical approach to balancing
  correctness, latency, and cost in massive-scale, unbounded, out-of-order data
  processing,'' \emph{Proceedings of the VLDB Endowment}, vol.~8, no.~12, pp.
  1792--1803, 2015.

\bibitem{akidau2013millwheel}
T.~Akidau, A.~Balikov, K.~Bekiro{\u{g}}lu, S.~Chernyak, J.~Haberman, R.~Lax,
  S.~McVeety, D.~Mills, P.~Nordstrom, and S.~Whittle, ``Millwheel:
  fault-tolerant stream processing at internet scale,'' \emph{Proceedings of
  the VLDB Endowment}, vol.~6, no.~11, pp. 1033--1044, 2013.

\bibitem{omission}
``{How NOT to Measure Latency},''
  https://www.infoq.com/presentations/latency-response-time, accessed:
  2017-07-11.

\bibitem{friedrich2017coordinated}
S.~Friedrich, W.~Wingerath, and N.~Ritter, ``Coordinated omission in nosql
  database benchmarking.'' in \emph{BTW (Workshops)}, 2017, pp. 215--225.

\end{thebibliography}

\end{document}